\documentclass[aps,pra,twocolumn]{revtex4-1}
\usepackage{amsmath,amssymb}
\usepackage{natbib}
\usepackage{braket}
\usepackage{hyperref}
\usepackage{graphicx}
\usepackage{color}
\usepackage[dvipsnames]{xcolor}
\usepackage[caption=false]{subfig}
\usepackage{epstopdf}
\usepackage{tikz}
\DeclareMathOperator{\Tr}{Tr}

\definecolor{byzantine}{rgb}{0.74, 0.2, 0.64}
\begin{document}
\title{Quantum Rabi Model with Two-Photon Relaxation}
\author{Moein Malekakhlagh}
\author{Alejandro W. Rodriguez}
\affiliation{Department of Electrical Engineering, Princeton University, New Jersey, 08544}
\begin{abstract}
We study a cavity-QED setup consisting of a two-level system coupled to a single cavity mode with two-photon relaxation. The system dynamics is modeled via a Lindblad master equation consisting of the Rabi Hamiltonian and a two-photon dissipator. We show that an even-photon relaxation preserves the $Z_2$ symmetry of the Rabi model, and provide a framework to study the corresponding non-Hermitian dynamics in the number-parity basis.  We discuss the role of different terms in the two-photon dissipator and show how one can extend existing results for the closed Rabi spectrum to the open case. Furthermore, we characterize the role of the $Z_2$ symmetry in the excitation-relaxation dynamics of the system as a function of light-matter coupling. Importantly, we observe that initial states with even-odd parity manifest qualitatively distinct transient and steady state behaviors, contrary to the Hermitian dynamics that is only sensitive to whether the initial state is parity-invariant. Moreover, the parity-sensitive dynamical behavior is not a creature of ultrastrong coupling and is present even at weak coupling values.
\end{abstract}
\maketitle

\textit{Introduction}. The Rabi model \cite{Rabi_Space_1937} describes the quantum interaction between a two-level system (TLS) and a bosonic mode. Despite its simple form, the Rabi model represents an important theoretical building block of quantized matter-field interactions and quantum information processing. It is applicable to a broad range of
quantum phenomena spanning microscopic to mesoscopic systems, finding realizations in a wide range of quantum platforms, including cavity-QED \cite{Brune_Quantum_1996, Raimond_Manipulating_2001, Mabuchi_Cavity_2002, Walther_Cavity_2006}, circuit-QED \cite{Wallraff_Strong_2004, Blais_Cavity_2004, Chiorescu_Coherent_2004, Schuster_Resolving_2007, Clarke_Superconducting_2008, Hofheinz_Synthesizing_2009}, nanoelectromechanical \cite{Irish_Quantum_2003, Cleland_Superconducting_2004, Schwab_Putting_2005, Lahaye_Nanomechanical_2009}, quantum-dot
\cite{Hennessy_Quantum_2007}, and trapped-ion \cite{Liebfried_Quantum_2003, Pedernales_Quantum_2015} systems.

Light-matter interactions within the Rabi model consist of rotating (resonant) and counter-rotating (non-resonant)
contributions. Traditionally, Rabi dynamics is analyzed under the rotating-wave approximation (RWA), resulting in the simplified Jaynes-Cummings (JC) model \cite{Jaynes_Comparison_1963}, valid when the coupling constant is much weaker than the TLS and mode frequencies. From the perspective of symmetry, RWA fictitiously extends the $Z_2$ symmetry of the model to a $U(1)$ symmetry, making the total excitation number the second conserved quantity besides the Hamiltonian and therefore facilitates analytical solutions. The JC model has been employed successfully to describe the dynamics of most cavity-QED setups \cite{Brune_Quantum_1996, Raimond_Manipulating_2001, Mabuchi_Cavity_2002, Walther_Cavity_2006}. However, with the advent of superconducting quantum devices, it has become feasible to reach
ultrastrong \cite{Niemczyk_Circuit_2010, Forn-Diaz_Observation_2010} and, more recently, deep-strong \cite{Yoshihara_Superconducting_2017} regimes of interactions. The breakdown of RWA in these regimes motivated various theoretical efforts to revise the Rabi model. First,
generalized versions of RWA \cite{Irish_Generalized_2007, Albert_Symmetric_2011} were introduced that captures correctly stronger couplings. Second, despite the contemporary understanding, Braak \cite{Braak_Integrability_2011} argued that the $Z_2$ symmetry of the Rabi model is sufficient for its integrability, showing that the regular spectrum in each parity subspace can be obtained from the roots of a transcendental function. Moreover, Chen \textit{et al.} provided a more physical derivation of the Rabi spectrum using Bugoliubov transformations \cite{Chen_Exact_2012}, contrary to the Bargmann representation \cite{Bargmann_Hilbert_1961} employed by Braak. These early studies paved the way toward ongoing developments of analytical
and perturbative methods for determining the spectrum, eigenmodes, and dynamics of the Rabi model under different parameter regimes \cite{Wolf_Exact_2012, Zhong_Analytical_2013, Dynamical_Wolf_2013, Rossatto_Spectral_2016, Xie_Quantum_2017}. 

\begin{figure}[t!]
\centering \includegraphics[scale=0.55]{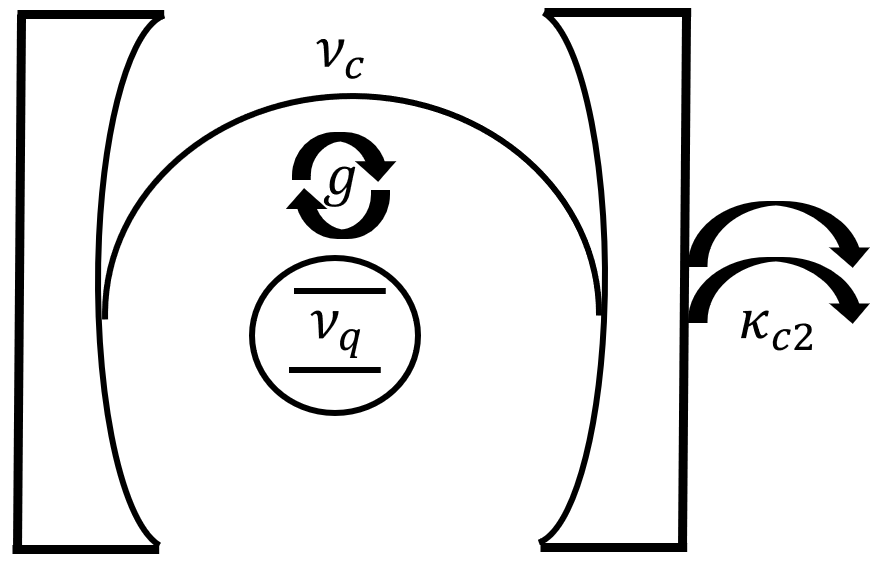}
\caption{Schematic of system consisting of a two-level system coupled to a single cavity mode with two-photon relaxation. We discuss possible physical realization of such a relaxation process in the Supplementary Material (SM), revisiting Refs.~\cite{Wolinsky_Quantum_1988, Leghtas_Confining_2015}}
\label{fig:SchematicSystem}
\end{figure}

To date, however, most works have focused primarily on ideal, closed (Hermitian) properties of the Rabi model, while role of $Z_2$ symmetry in realistic, open (non-Hermitian) scenarios remains an open question. An even exchange of excitations between a cavity mode and environment conserves the $Z_2$ symmetry. The latter is particularly important given emerging studies of the dynamics of a single cavity mode under two-photon relaxation \cite{Simaan_Quantum_1975, Simaan_Off-diagonal_1978, Voigt_Exact_1980, Gea_Two-photon_1989, Gilles_Two-Photon_1993, Klimov_Algebraic_2003, Voje_Multi-Photon_2013, Albert_Symmetries_2014}. Such a relaxation process has been recently implemented in circuit-QED \cite{Leghtas_Confining_2015} following a four-wave mixing scheme proposed first by Wolinsky and Carmichael \cite{Wolinsky_Quantum_1988} (see SM \footnote{The supplementary material provides further discussion on physical realization of two-photon relaxation, $Z_2$ symmetry of the Rabi model and comparison between phenomenological and full spectrum. It includes the following references: \cite{Wolinsky_Quantum_1988, Leghtas_Confining_2015, Felicetti_Two-Photon_2018, Casanova_Deep_2010, Yi_Effective_2001, Prosen_Spectral_2010, Kolda_Tensor_2009}}). A major motivation behind recent studies of even-photon relaxation processes is their application to realization of dynamically protected, universal quantum computing paradigms
\cite{Mirrahimi_Dynamically_2014, Sun_Tracking_2014, Ofek_Extending_2016, Wang_Schrodinger_2016}, in which the quantum information is encoded in logical qubits consisting of Schr\"odinger cat states with distinct parity that exhibit reliable protection to photon dephasing and single-photon relaxation errors \cite{Mirrahimi_Dynamically_2014}.

In this Letter, we generalize the theory of $Z_2$ symmetry of the Rabi model to the open quantum case. We first review the spectrum of the closed Rabi Hamiltonian, providing analytical recursion relations for both the eigenfrequency and eigenmodes of the system. Our analysis and calculation are performed in the (cavity) number-(overall) parity representation
\cite{Casanova_Deep_2010}, where the $Z_2$ symmetry of the model is explicit. For the open scenario, we consider a Lindblad master equation \cite{Lindblad_Generators_1976,
  Gorini_Completely_1976} of the Rabi Hamiltonian with two-photon dissipation for the cavity mode. To analyze its spectral properties, we employ an effective Hamiltonian obtained by keeping diagonal decay terms, while neglecting the off-diagonal collapse in the two-photon dissipator. This phenomenological treatment provides a reliable approximation to the complex eigenfrequencies, but not necessarily the eigenmodes and ground state. While an exact definition of a full effective Hamiltonian exists, mapping the Lindblad dynamics into a norm-preserving Schrodinger equation~\cite{Yi_Effective_2001}, analytical treatment of its associated spectrum seems prohibitive due to its significantly larger Hilbert space compared to the phenomenological model (see SM for comparison). We follow numerical integration of the Lindblad equation for studying the dynamics, while the effective phenomenological Hamiltonian is primarily used for approximate analytical discussion of the spectrum and a better understanding of the observed dynamics.


\textit{Model}. Our system consists of a TLS coupled to a single cavity mode, engineered such that single-photon is negligible compared to two-photon relaxation, constraining it to exchange only pairs of photons with the environment (Fig.~\ref{fig:SchematicSystem}). We model the system dynamics via the Linbdlad equation
\begin{subequations}
\begin{align}
&\dot{\hat{\rho}}(t)=-i[\hat{H}_s,\hat{\rho}(t)]+2\kappa_{c2}\mathcal{D}[\hat{a}^2]\hat{\rho}(t),
\label{eqn:Model-Lindblad Eq 1}\\
&\hat{H}_{\text{s}}\equiv \nu_c\hat{a}^{\dag}\hat{a}+\frac{\nu_q}{2}\hat{\sigma}^z+g\left(\hat{a}+\hat{a}^{\dag}\right)\left(\hat{\sigma}^-+\hat{\sigma}^+\right),
\label{eqn:Model-Hs}
\end{align}
\end{subequations}
with $\nu_q$, $\nu_c$, and $g$ denoting the qubit frequency, cavity frequency, and light-matter coupling, respectively. Two-photon relaxation is described via the dissipator, $\mathcal{D}[\hat{a}^2](\bullet)=\hat{a}^2(\bullet)(\hat{a}^{\dag})^2-\frac{1}{2}\left\{(\hat{a}^{\dag})^2\hat{a}^2,(\bullet)\right\}$, with $\kappa_{c2}$ denoting the two-photon relaxation rate.
 
We next transform the Lindblad Eq.~(\ref{eqn:Model-Lindblad Eq 1}) such that the $Z_2$ symmetry of the Rabi Hamiltonian and two-photon relaxation become explicit (see SM). In particular, we define the overall parity operator for the system as
\begin{align}
\hat{P}=\hat{P}_q\hat{P}_c=e^{i\pi\hat{\sigma}^+\hat{\sigma}^-}e^{i\pi\hat{a}^{\dag}\hat{a}}=-\hat{\sigma}^{z}e^{i\pi\hat{a}^{\dag}\hat{a}}.
\label{eqn:Model-Def of ParOp}
\end{align}
The $Z_2$ symmetry of the Rabi Hamiltonian~(\ref{eqn:Model-Hs}) means that $\hat{P}^{\dag}\hat{H}_{s}\hat{P}=\hat{H}_s$. Consequently, the Hilbert space can be partitioned into parity subspaces having even (plus) and odd (minus) total excitation numbers:
\begin{subequations}
\begin{align}
&p=+1: \ \{\ket{0,g}, \ket{1,e},\ket{2,g}, \ket{3,e},\ket{4,g}, \ldots\},
\label{eqn-RabiSpec-P=+1 subspace}\\
&p=-1: \ \{\ket{0,e}, \ket{1,g},\ket{2,e}, \ket{3,g},\ket{4,e}, \ldots\}.
\label{eqn-RabiSpec-p=-1 subspace}
\end{align}
\end{subequations}
The adjacent states in each subspace are coupled via both rotating or the counter-rotating terms. If we neglect the latter, each subspace is reduced into a collection of number-conserving Jaynes-Cummings doublets $\{\ket{n-1,e},\ket{n,g}\}$, given by:
\begin{subequations}
\begin{align}
&p=+1: \ \{\ket{0,g}\}, \{\ket{1,e},\ket{2,g}\}, \{\ket{3,e}, \ket{4,g}\}, \ldots,
\label{eqn:Model-P=+1 doublets}\\
&p=-1: \ \{\ket{0,e}, \ket{1,g}\},\{\ket{2,e}, \ket{3,g}\},\ldots .
\label{eqn:Model-P=-1 doublets}
\end{align}
\end{subequations} 

\begin{table}
\begin{tabular}{|c|c|c|}
\hline
\textbf{number-excitation basis} & $\ket{n,g}$ & $\ket{n,e}$ \\ \hline
\textbf{number-parity basis} & $\ket{n,(-1)^{n}}$ & $\ket{n,(-1)^{n+1}}$ \\ \hline
\end{tabular}
\caption{Correspondence between (cavity) number-(qubit) excitation and (cavity) number- (overall) parity bases.}
\label{tab:BasesCorrespondence}
\end{table}

Defining a new set of bosonic operators, $\hat{b}\equiv\hat{\sigma}^x \hat{a}$, and replacing $\hat{\sigma}^{z}$ in terms of the parity operator of Eq.~(\ref{eqn:Model-Def of ParOp}), one can rewrite the Rabi Hamiltonian~(\ref{eqn:Model-Hs}) as \cite{Casanova_Deep_2010}
\begin{align}
\hat{H}_{\text{s}}=\nu_c\hat{b}^{\dag}\hat{b}-\frac{\nu_q}{2}e^{i\pi\hat{b}^{\dag}\hat{b}}\hat{P}+g\left(\hat{b}+\hat{b}^{\dag}\right).
\label{eqn:Model-Hs 2}
\end{align}
The parity $\hat{P}$ and bosonic $\hat{b}$ operators commute, and thus provide a complete basis for the Hilbert space defined as $\hat{b}^{\dag}\hat{b}\ket{n,p}=n\ket{n,p}$ and $\hat{P}\ket{n,p}=p\ket{n,p}$ for $n=0, 1, 2, \ldots$ and $p=\pm 1$, respectively. Table~\ref{tab:BasesCorrespondence} summarizes the correspondence between the (old) number-excitation and (new) number-parity bases.

Next, we rewrite the original Lindblad Eq.~(\ref{eqn:Model-Lindblad Eq 1}) in this basis, starting by observing that the two-photon dissipator is also invariant under the parity transformation, i.e. $\hat{P}^{\dag}\mathcal{D}[\hat{a}^2]\hat{P}=\mathcal{D}[(-\hat{a})^2]=\mathcal{D}[\hat{a}^2]$, where $\hat{a}^2=(\hat{\sigma}^x\hat{a})^2=\hat{b}^2$ also implies that $\mathcal{D}[\hat{a}^2]=\mathcal{D}[\hat{b}^2]$. In the quantum treatment of dissipation, the two contributions to the dissipator are described by decay and collapse terms.  The former represents the rate at which a quantum state loses probability while the latter represents the rate at which lower states in the excitation ladder receive probability, in such a way that the net probability is conserved in time, i.e. $\Tr\left(\mathcal{D}[\hat{b}^2]\hat{\rho}\right)=0$. Separating the two contributions, one can reexpress the Lindblad Eq.~(\ref{eqn:Model-Lindblad Eq 1}) to yield,
\begin{subequations}
\begin{align}
\dot{\hat{\rho}}(t)=-i\left[\hat{H}_{\text{s,ef}}\hat{\rho}(t)-\hat{\rho}(t)\hat{H}_{\text{s,ef}}^{\dag}\right]+2\kappa_{c2}\hat{b}^2\hat{\rho}(t)(\hat{b}^{\dag})^2,
\label{eqn:Model-Lindblad Eq 2}
\end{align}
with $\hat{H}_{s,\text{ef}}$ denoting the phenomenological effective Hamiltonian as
\begin{align}
\hat{H}_{\text{s,ef}}=\nu_c\hat{b}^{\dag}\hat{b}-\frac{\nu_q}{2}e^{i\pi\hat{b}^{\dag}\hat{b}}\hat{P}+g\left(\hat{b}+\hat{b}^{\dag}\right)-i\kappa_{c2}(\hat{b}^{\dag})^2\hat{b}^2.
\label{eqn:Model-Def of H_s,ef}
\end{align}
\end{subequations}
Neglecting the coupling induced by collapse, the last term in
Eq.~(\ref{eqn:Model-Lindblad Eq 2}), the dissipative dynamics is
approximated by $\hat{H}_{\text{s,ef}}$. This framework is a middle ground in which the unitary part of the system dynamics is treated quantum mechanically, while the dissipation is treated phenomenologically. Essentially, such an approach provides a good approximation for the complex spectrum of the problem, while ignoring proper characterization of the modal and ground state information (See Sec.~IV of the SM for further discussion). 


\textit{Spectrum}. Here, we first revisit the spectrum of the closed Rabi model and benchmark our solution against those of Braak \cite{Braak_Integrability_2011}. For the open case, we study the impact of two-photon relaxation via $\hat{H}_{\text{s,ef}}$ of Eq.~(\ref{eqn:Model-Def of H_s,ef}). In particular, we show that the typical solution obtained for the closed case can be generalized to yield the complex eigenfrequencies of the open system.

\begin{figure}[t!]
\centering
\subfloat[\label{subfig:PhenOpRabiSpecVq8Em1Vc1Qc40FuncOfg}]{%
\includegraphics[scale=0.32]{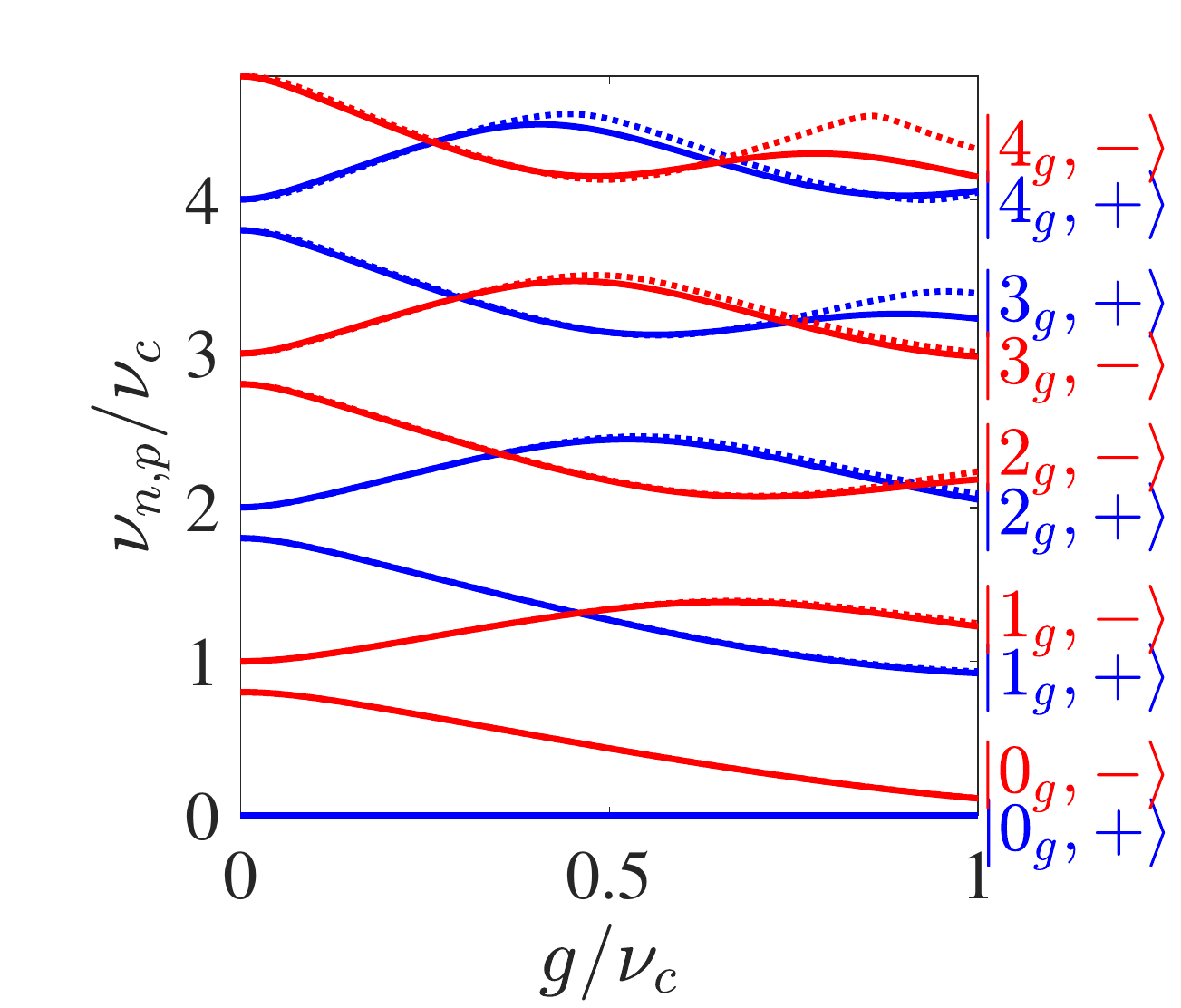}%
}
\subfloat[\label{subfig:PhenOpRabiDissVq8Em1Vc1Qc40FuncOfg}]{%
\includegraphics[scale=0.32]{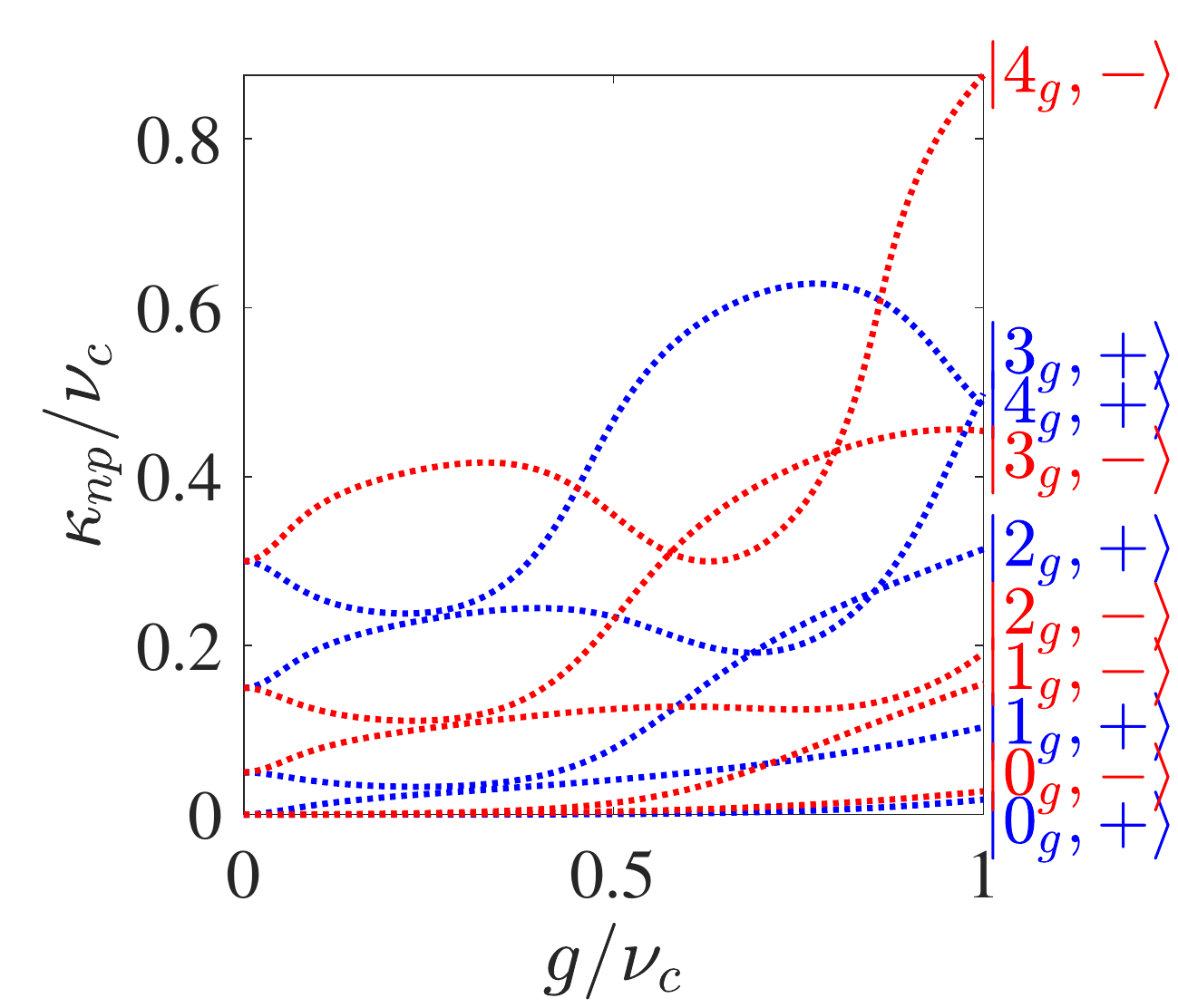}%
}\\
\subfloat[\label{subfig:PhenOpRabiComplexSpecVq8Em1Vc1Qc40FuncOfg}]{%
\includegraphics[scale=0.27]{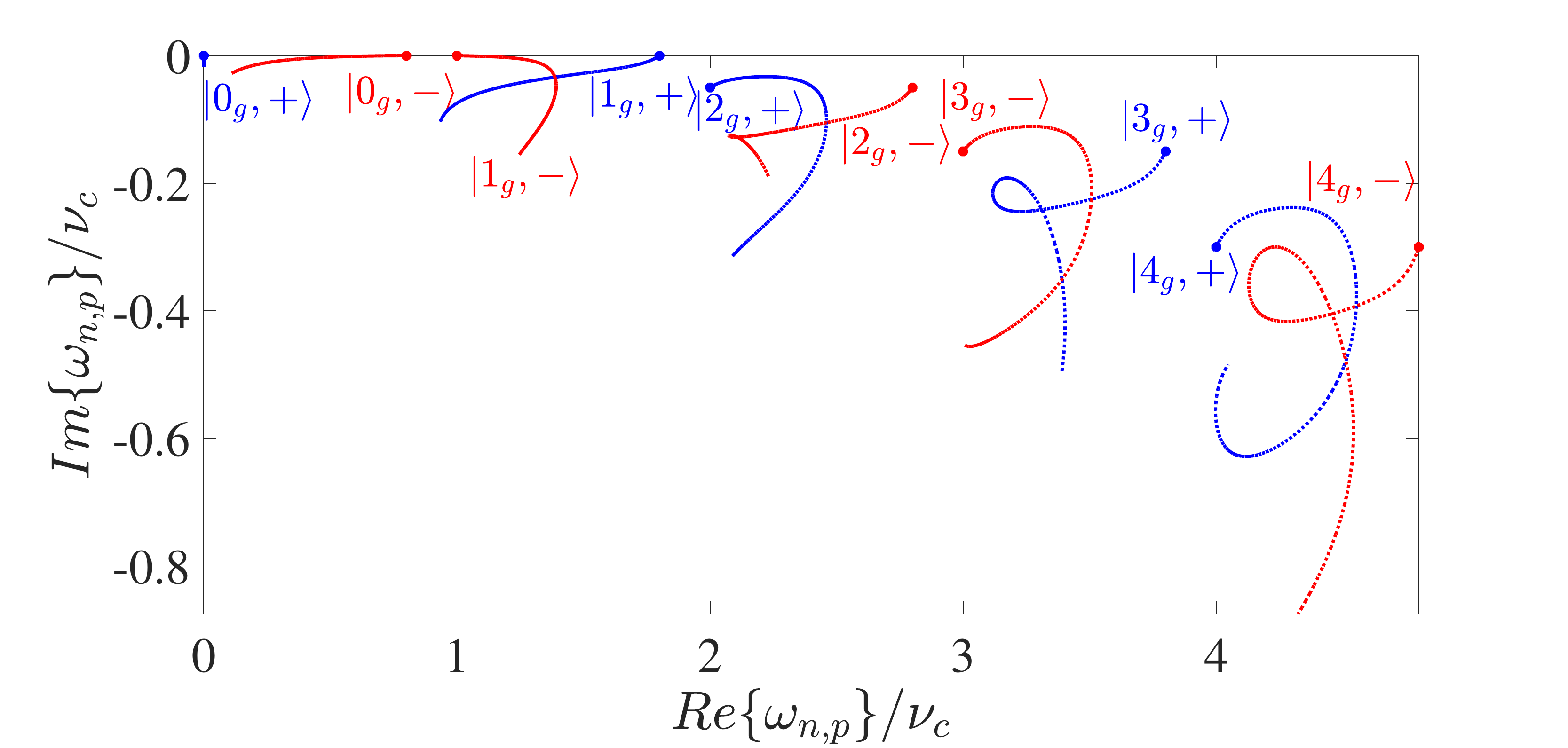}%
}
\caption{Phenomenological open Rabi eigenfrequencies
  $\omega_{n,p}\equiv \nu_{n,p}-i\kappa_{n,p}$ for $\nu_q=0.8\nu_c$ and $\kappa_{c2}=\nu_c/40$.  a) Real part (frequency), b) imaginary part (decay rate), and c) complex spectrum as a function of light-matter coupling $g$. Solid lines in a) show the result for the closed ($\kappa_{c2}=0$) case while dotted lines are for $\kappa_{c2}=\nu_c/40$. The labels $\ket{n_g,\pm}$ in a) and b) are ordered based on values at $g=0$. The frequencies in a) are plotted relative to the ground state $\ket{0_g,+}$.}
\label{fig:PhenOpRabiSpec} 
\end{figure}

We begin with the eigenvalue problem for the closed Rabi model, $\hat{H}_{\text{s,p}}\ket{n_g,p}=\omega_{np}\ket{n_g,p}$, where $n_g$ labels the eigenvalues and eigenmodes at a nonzero $g$ and $p$ is the corresponding parity subspace. Expanding the unknown eigenmodes in terms of the number-parity basis, $\ket{n_g,p}=\sum\limits_{m=0}^{\infty} c_{np,m}\ket{m,p}$, one finds that the eigenfrequencies $\omega_{np}$ are obtained by the roots $G_{p}(\omega_{np})=0$, where $G_p\equiv \lim\limits_{m\to\infty} G_{p,m}$ and $G_{p,m}$ satisfies the following recursion relation (see SM):
\begin{align}
G_{p,m}=\alpha_{np,m} G_{p,m-1}-\beta_{p,m-1}\gamma_{p,m}G_{p,m-2},
\label{eqn:RabiSpec-RecRel G_p,m}
\end{align}
subject to initial conditions, $G_{p,0}=\alpha_{p,0}$ and
$G_{p,1}=\alpha_{p,0} \alpha_{p,1}-\beta_{p,0}\gamma_{p,1}$. The coefficients in the recursion Eq.~(\ref{eqn:RabiSpec-RecRel G_p,m}) read
\begin{align}
\begin{split}
&\alpha_{np,m}\equiv \omega_{np}-m\nu_c+\frac{p}{2}(-1)^m\nu_q,
\\
&\beta_{p,m}\equiv-\sqrt{m+1}g, \ \gamma_{p,m}\equiv-\sqrt{m}g.
\end{split}
\label{eqn:RabiSpec-Def of alpha_p,m beta_p,m gamma_p,m}
\end{align}
Similarly, the corresponding eigenmodes are determined by yet another recursion relation for the probability amplitudes $c_{np,m}$, given by:
\begin{align}
\begin{split}
\alpha_{np,m}c_{np,m}&+\beta_{p,m}c_{np,m+1}+\gamma_{p,m}c_{np,m-1}=0,
\end{split}
\label{eqn:RabiSpec-RecRel c_p,m}
\end{align}
with initial conditions, $\alpha_{p,0}c_{p,0}+\beta_{p,0}c_{p,1}=0$. An illustrative example of the variation of the spectrum with respect to $g$ is shown in Fig.~(\ref{subfig:PhenOpRabiSpecVq8Em1Vc1Qc40FuncOfg}), with
parameters chosen to compare our results with those in Fig.~2 of Ref.~\cite{Braak_Integrability_2011}.


Within the phenomenological treatment of relaxation, the system
dynamics are determined by $\hat{H}_{\text{s,ef}}$ of Eq.~(\ref{eqn:Model-Def of H_s,ef}). Here, we find that the recursion relations determining the eigenfrequencies~(\ref{eqn:RabiSpec-RecRel G_p,m}) and eigenmodes~(\ref{eqn:RabiSpec-RecRel c_p,m}) have the same form as those of the closed system, except that the coefficients $\alpha_{np,m}$ are modified as $\alpha_{np,m}\rightarrow \alpha_{np,m}+im \left(m-1\right)\kappa_{c2}$ (see SM). To understand the changes induced by phenomenological decay (compared to the closed), we first consider the regime of zero coupling $g=0$, where the decay terms are diagonal in the number basis. In this scenario, the $m$th bare cavity mode acquires a decay rate of $\kappa_{c2}m (m-1)$, resulting in nonzero values for all cavity number states except the ground and first-excited state, for each parity. As the coupling $g$ is turned on, the hybridization between the qubit and the cavity mode allows these terms not only to induce additional decay, but also modify the real frequency of each state. Figure~\ref{fig:PhenOpRabiSpec} (dotted lines) shows such hybridization as a function of $g$, as calculated by the phenomenological model. We note that an analogous phenomenological model based on the JC model can be solved analytically and result in decay rates that plateau at ultrastrong coupling values of $g$ and hence mischaracterizes the interplay of light-matter coupling and two-photon relaxation (see SM for comparison).

\textit{Excitation-relaxation dynamics}. Here, we study the
dissipative dynamics of the system and discuss the role of
$Z_2$ symmetry. For concreteness, we consider the situation in which the cavity is initially prepared with an even or odd number of photons, and describe the ensuing dynamics of the cavity photon and qubit population as a function of both time and $g$. In particular, we consider two scenarios of starting with two [$\hat{\rho}(0)=\ket{2,g}\bra{2,g}$] or three [$\hat{\rho}(0)=\ket{3,g}\bra{3,g}$] initial cavity photons and the qubit in ground state, as representatives of the plus or minus parity subspaces. Due to pair-exchange of photons with the environment, we intuitively expect states with even or odd initial cavity photons to exhibit different transient and steady state behavior.

First, consider the simplest case of $g=0$. This choice of parameter decouples the qubit and hence corresponds to the problem of a single cavity mode with two-photon relaxation, which has been studied in detail using multiple methods \cite{Gea_Two-photon_1989, Gilles_Two-Photon_1993, Klimov_Algebraic_2003, Voje_Multi-Photon_2013, Albert_Symmetries_2014}. In this case, initial states having even (odd) numbers of cavity photons end up with zero (one) cavity photons in the steady state \footnote{Note that states with mixed parity can always be expressed as linear combinations of two independent problems.}.
\begin{figure}[t!]
\centering
\subfloat[\label{subfig:ExRe-AdAVcPiQc40ICTwoPhFuncOfg}]{%
\includegraphics[scale=0.31]{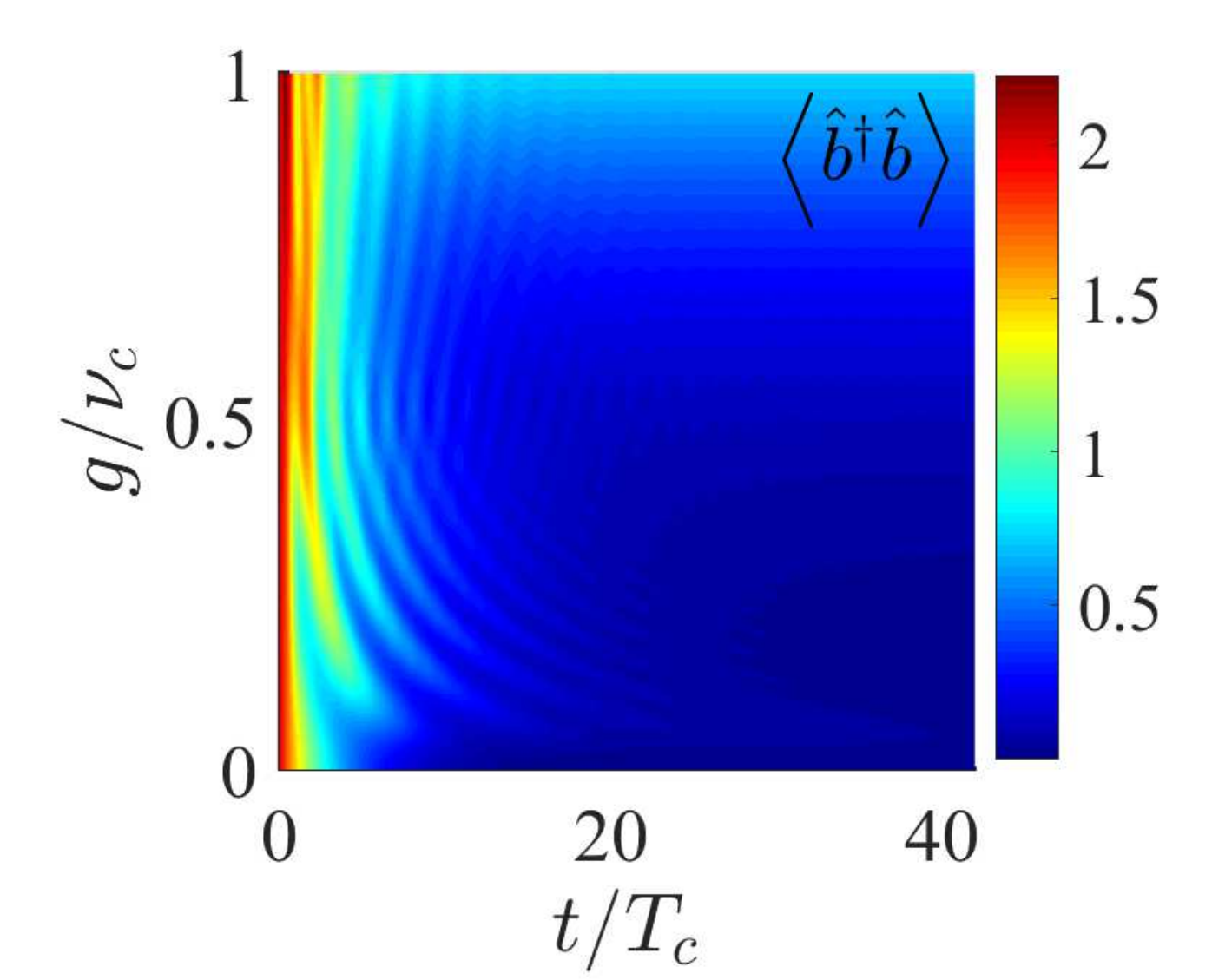}%
} 
\subfloat[\label{subfig:ExRe-SpSmVcPiQc40ICTwoPhFuncOfg}]{%
\includegraphics[scale=0.31]{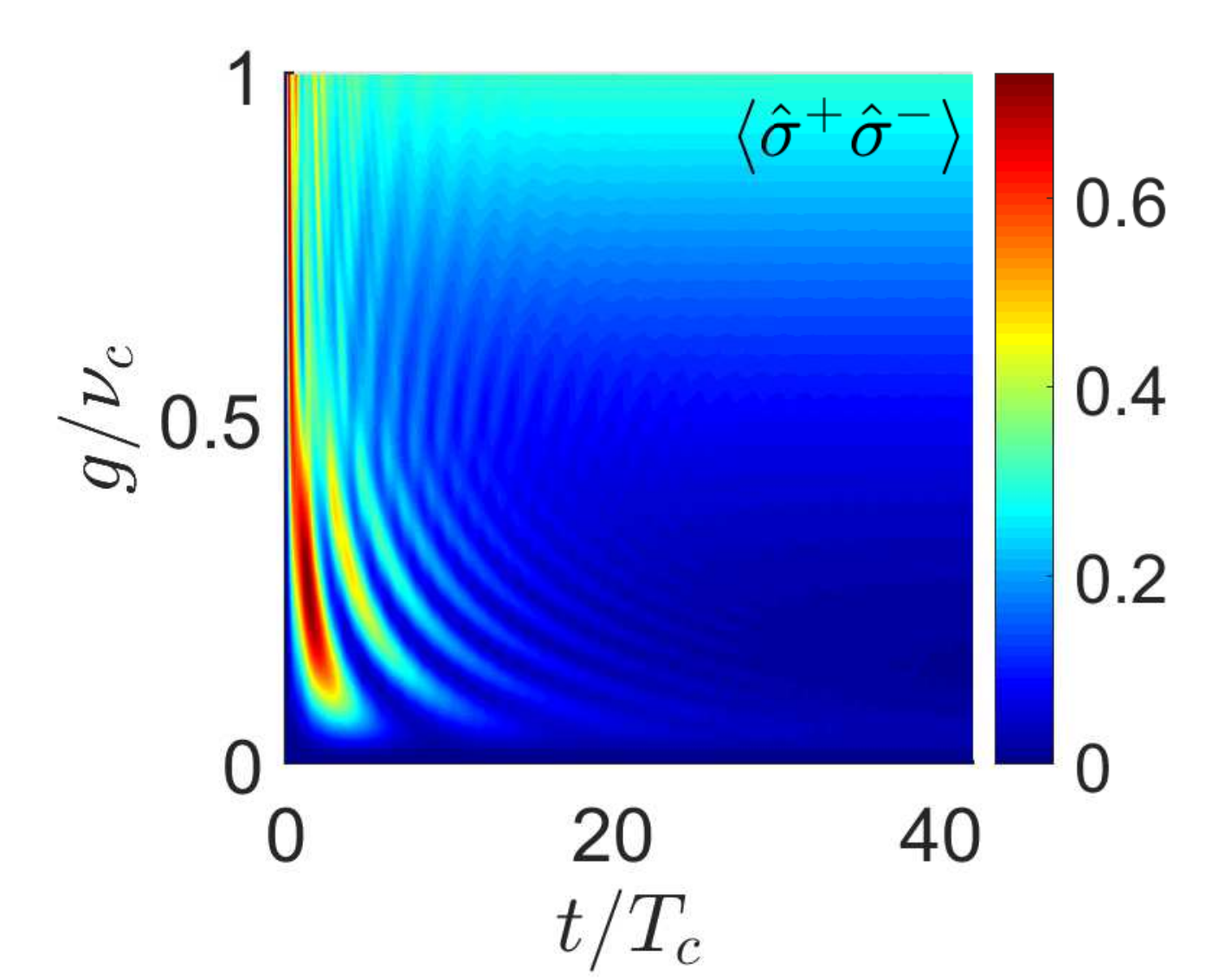}%
}\\
\subfloat[\label{subfig:ExRe-IcDistVcPiQc40ICTwoPhFuncOfg}]{%
\includegraphics[scale=0.115]{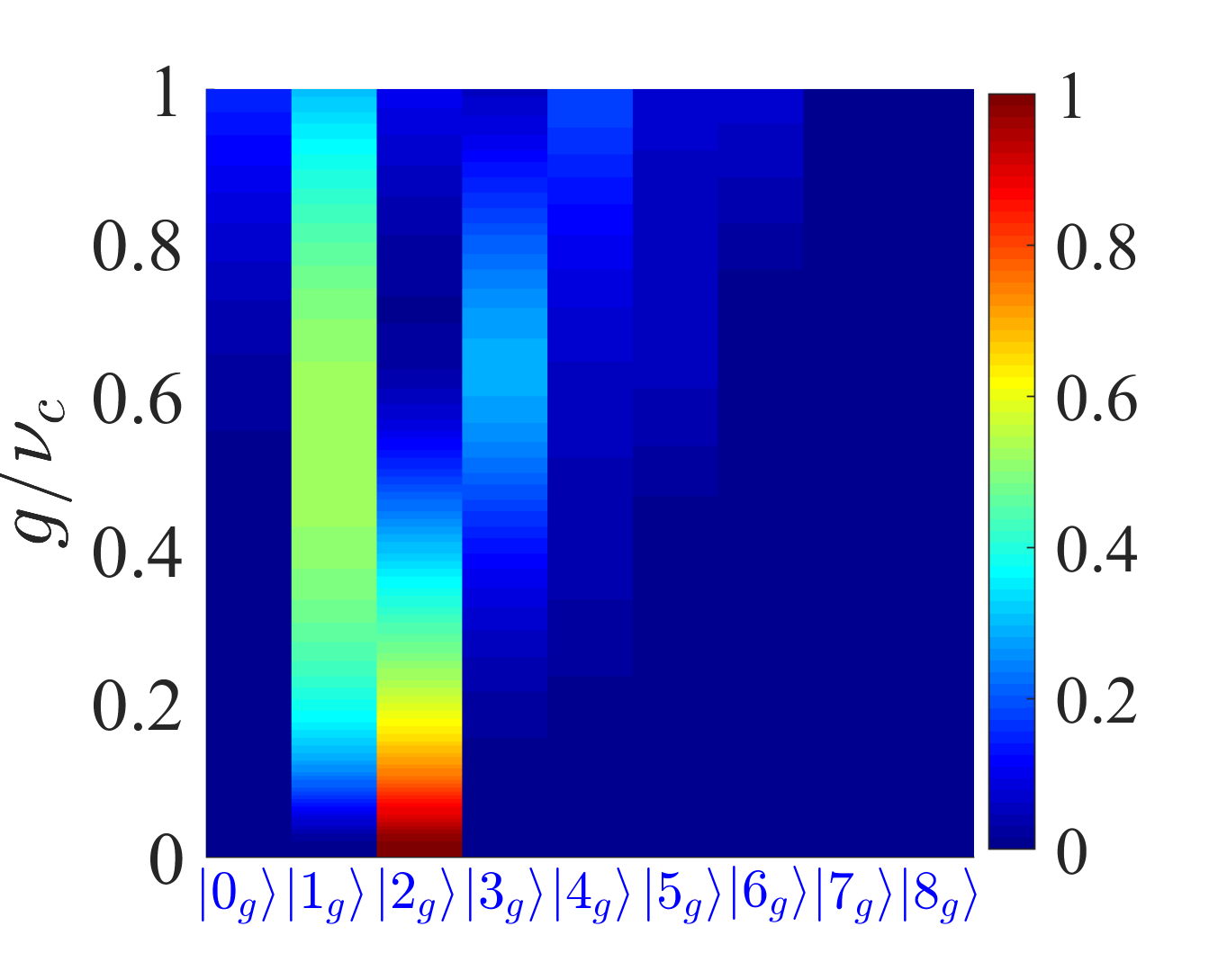}%
}
\subfloat[\label{subfig:ExRe-SSVcPiQc40ICTwoPhFuncOfg}]{%
\includegraphics[scale=0.30]{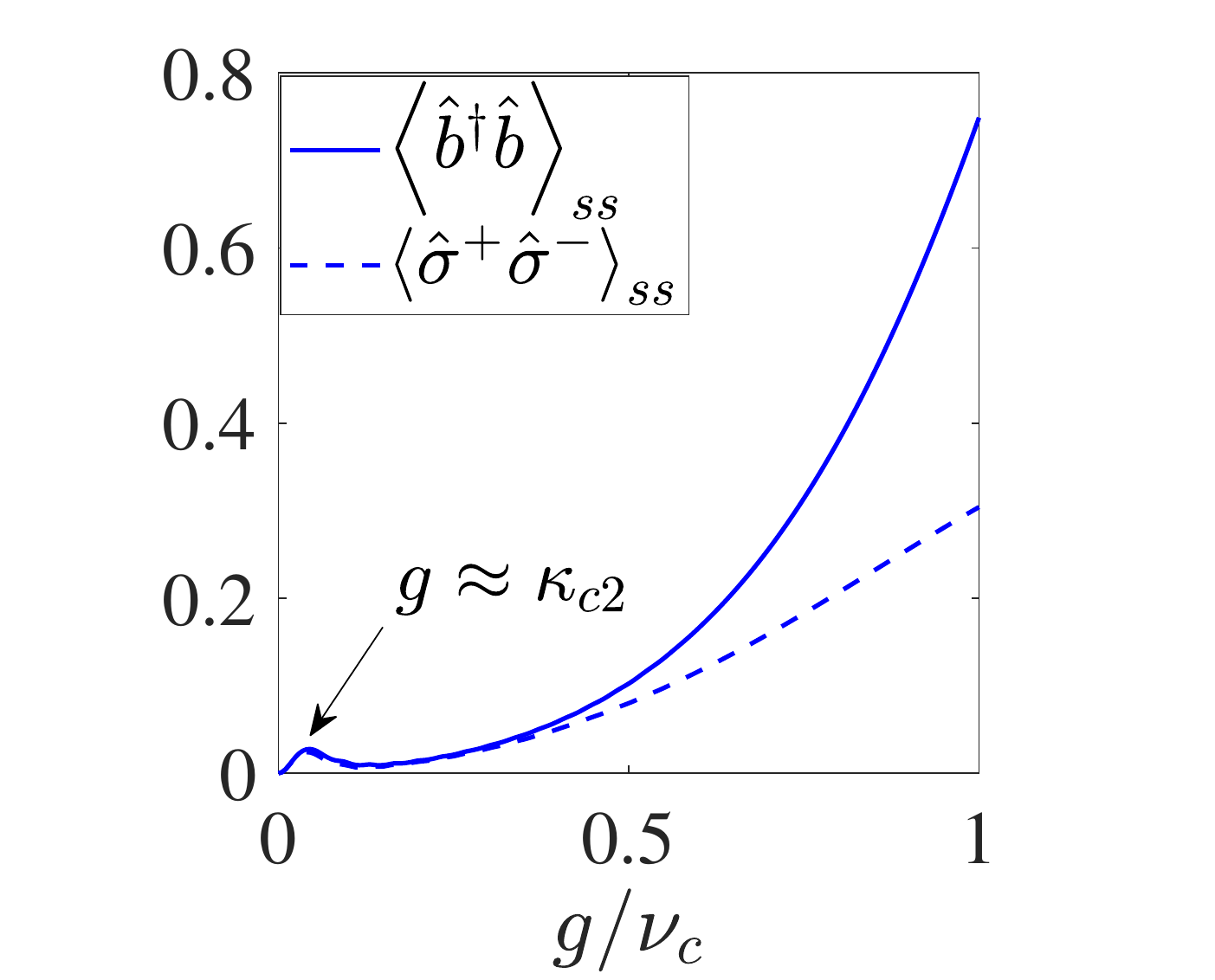}%
}
\caption{Excitation-relaxation dynamics of the system of Fig.~(2) when the system is prepared with \textcolor{blue}{two} cavity photons and the qubit is in the ground state, i.e. $\hat{\rho}(0)=\ket{2,g}\bra{2,g}=\ket{2,+}\bra{2,+}$, as a function of light-matter coupling $g$. a) Cavity photon number, b) qubit excitation number, and c) mapping of the bare state $\color{blue}\ket{2,g}\color{black}$ to the eigenmodes in the \textcolor{blue}{even (+)} parity subspace. For convenience, we omit the parity index in the $x$-axis. d) Steady state populations. Model parameters are the same as in Fig.~\ref{fig:PhenOpRabiSpec}. The time axis in a) and b) is normalized to half of the cavity round-trip time $T_c\equiv \pi/\nu_c$. The two-photon relaxation time reads $T_{\kappa 2}\equiv 1/\kappa_{c2}=40T_c/\pi$. The cavity mode Hilbert space cut	off is chosen as $N_c=9$.}
\label{fig:ExRe-ICTwoPhFuncOfg} 
\end{figure}
\begin{figure}[t!]
\centering
\subfloat[\label{subfig:ExRe-AdAVcPiQc40ICThreePhFuncOfg}]{%
\includegraphics[scale=0.31]{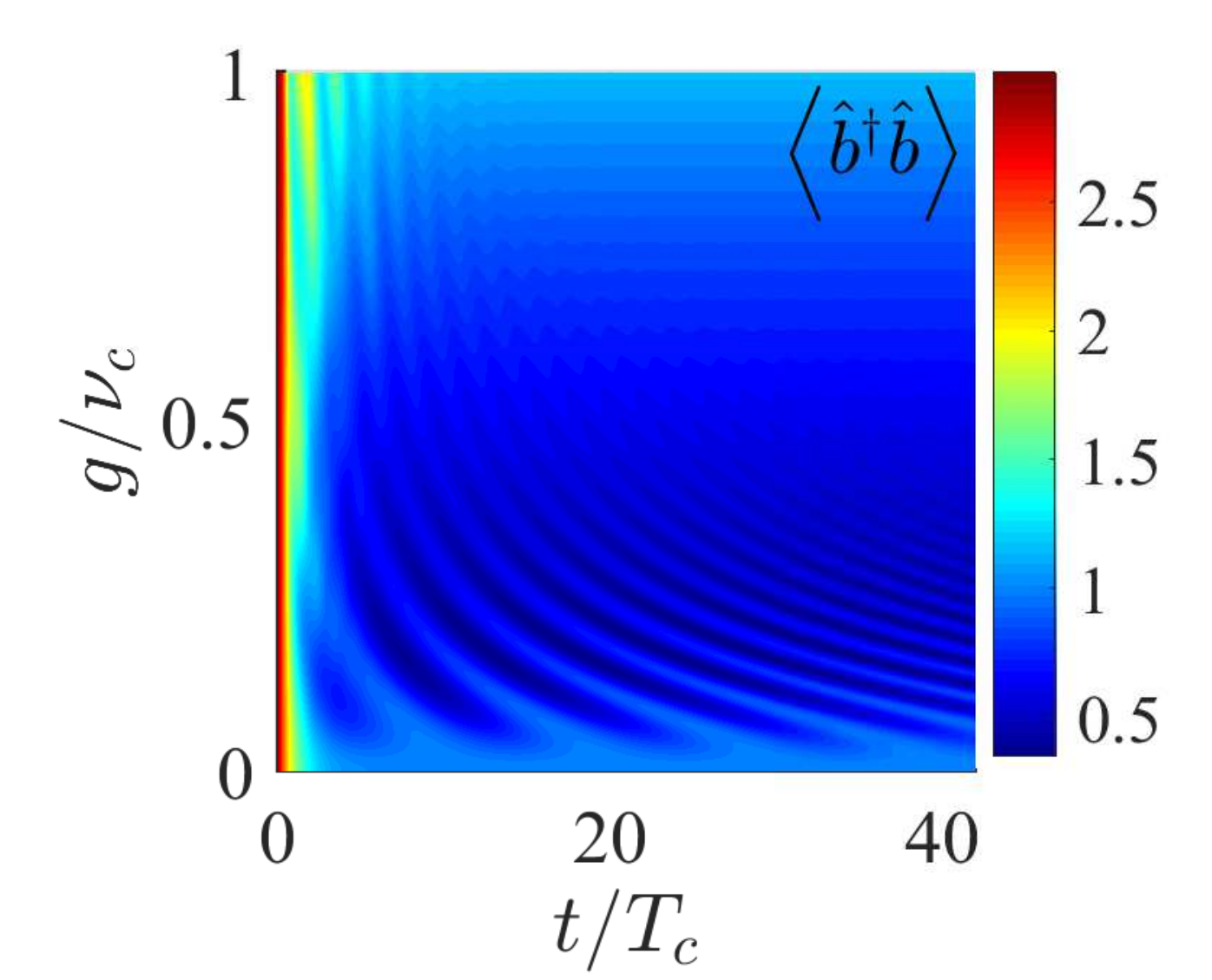}%
}
\subfloat[\label{subfig:ExRe-SpSmVcPiQc40ICThreePhFuncOfg}]{%
\includegraphics[scale=0.31]{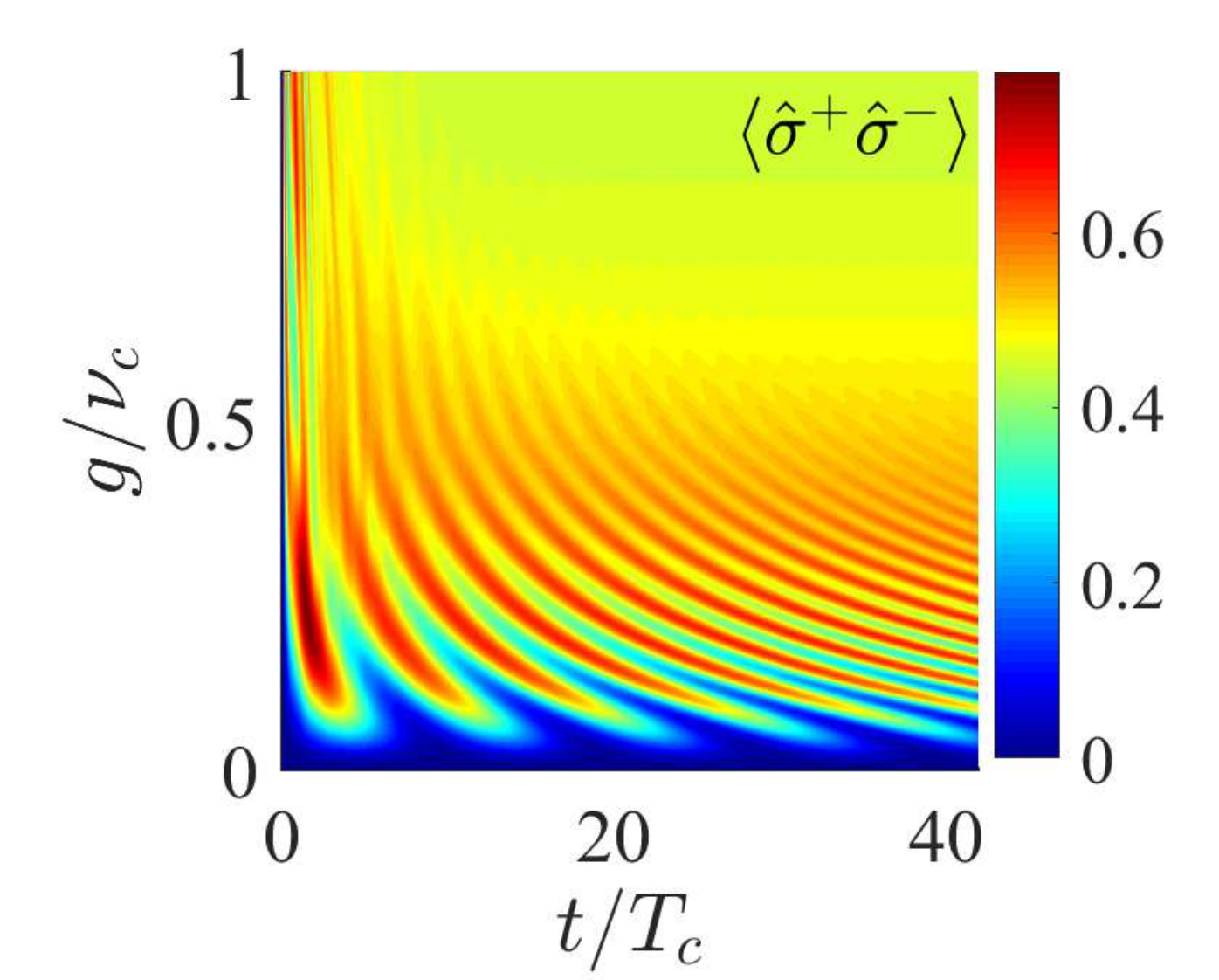}%
}\\
\subfloat[\label{subfig:ExRe-IcDistVcPiQc40ICThreePhFuncOfg}]{%
\includegraphics[scale=0.115]{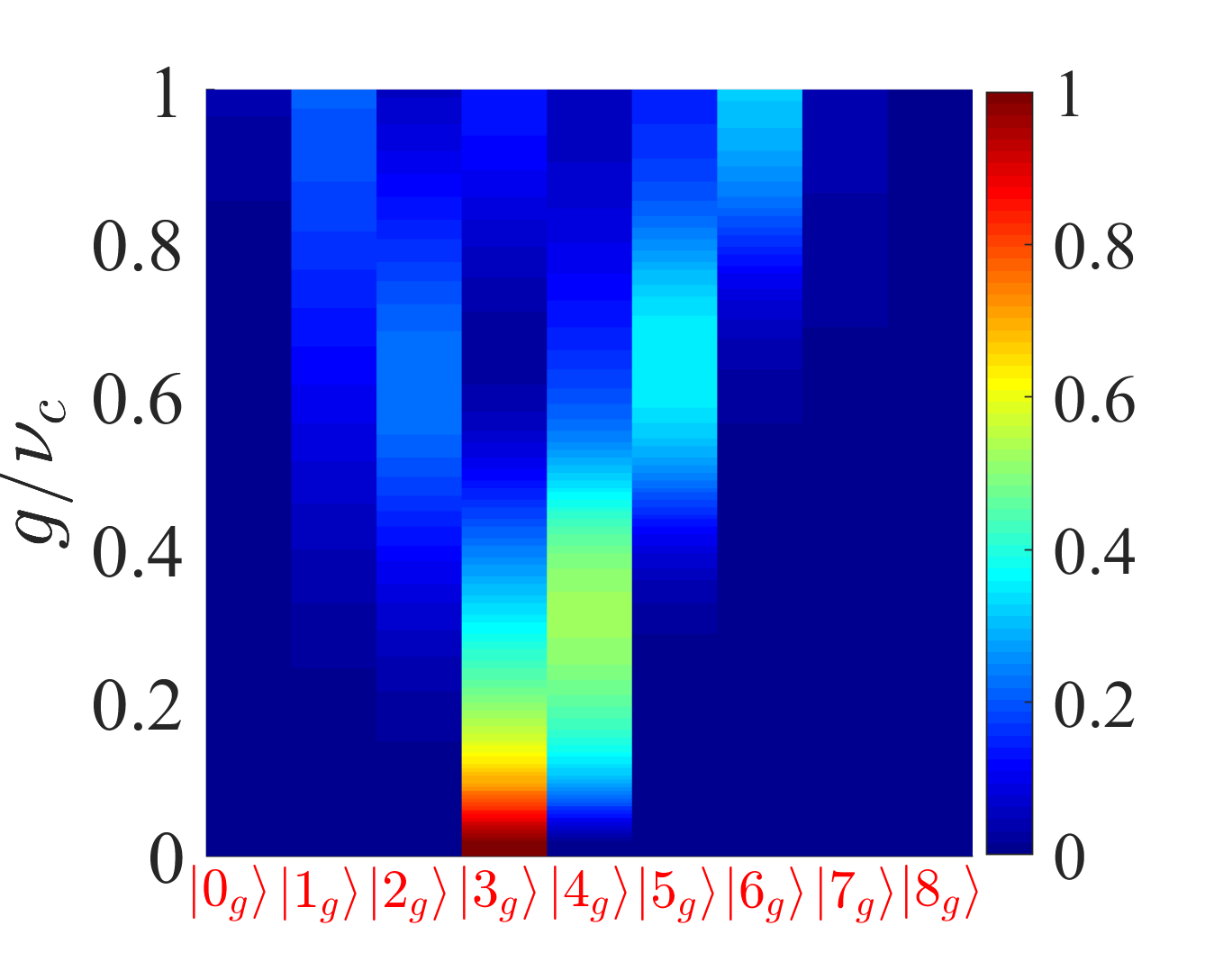}%
} 
\subfloat[\label{subfig:ExRe-SSVcPiQc40ICThreePhFuncOfg}]{%
\includegraphics[scale=0.30]{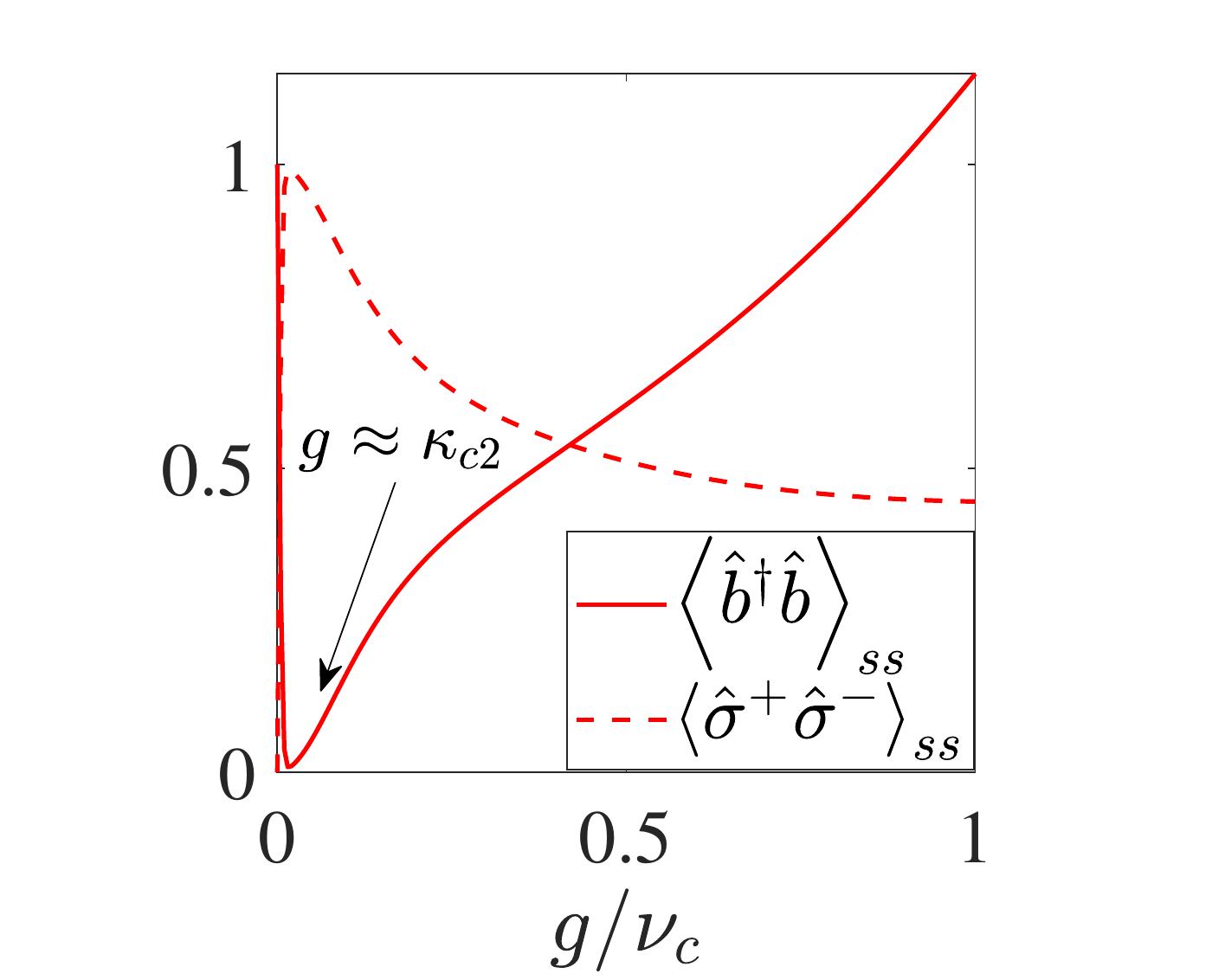}%
}
\caption{Excitation-relaxation dynamics when the system is prepared with \textcolor{red}{three} cavity photons and the qubit is in the ground state, i.e. $\hat{\rho}(0)=\ket{3,g}\bra{3,g}=\ket{3,-}\bra{3,-}$, as a function of light-matter coupling $g$. The figure follows the same format as Fig.~(3), except that the bare state $\color{red}\ket{3,g}\color{black}$ is instead mapped to eigenmodes in the \textcolor{red}{odd (-)} parity subspace. Other parameters are the same as in Fig.~\ref{fig:ExRe-ICTwoPhFuncOfg}.}
\label{fig:ExRe-ICThreePhFuncOfg} 
\end{figure}

Next, we move on to characterize the interplay of two-photon
relaxation and the qubit for $g \neq 0$. Here, closed form analytical solutions of the evolution operator at arbitrary $g$ seem intractable, and instead we employ numerical integration of the Lindblad Eq.~(\ref{eqn:Model-Lindblad Eq 2}). The time evolution of the cavity/qubit excitations as a function of $g$ is studied in Figs.~\ref{fig:ExRe-ICTwoPhFuncOfg}
and~\ref{fig:ExRe-ICThreePhFuncOfg} for the cases of two and three initial cavity photons, correspondingly. In both cases, it is generally observed that as $g$ is increased, more complex beatings between various normal modes emerge. Such beatings can be approximately understood from the mapping of the initial cavity state to the corresponding eigenmodes of the open Rabi model. This shows which modes are more active at a given value of $g$ in each parity subspace (Figs.~\ref{subfig:ExRe-IcDistVcPiQc40ICTwoPhFuncOfg} and \ref{subfig:ExRe-IcDistVcPiQc40ICThreePhFuncOfg}). For example, for the case of $\hat{\rho}(0)=\ket{2,g}\bra{2,g}$, the initial probability is shared between states $\ket{1_g,+}$ and $\ket{2_g,+}$ up to intermediate values of g ($0 < g \lesssim 0.5 \nu_c$), beyond which $\ket{1_g,+}$ and $\ket{3_g,+}$ dominate. The corresponding frequency and decay rate of the modes can be obtained from
Figs.~\ref{subfig:PhenOpRabiSpecVq8Em1Vc1Qc40FuncOfg}-\ref{subfig:PhenOpRabiDissVq8Em1Vc1Qc40FuncOfg}.

Despite this generic similarity, it is observed that
due to the nontrivial interplay of light-matter coupling and
two-photon relaxation, the two cases under consideration have
different transient and steady state characteristics. For the case of two initial cavity photons, we observe that the system reaches steady state on a timescale that is more or less given by the two-photon relaxation rate $\kappa_{c2}$
(Figs.~\ref{subfig:ExRe-AdAVcPiQc40ICTwoPhFuncOfg}-\ref{subfig:ExRe-SpSmVcPiQc40ICTwoPhFuncOfg}). On
the other hand, in the case of three initial cavity photons, the
transient dynamics has more features. Generally, at small $g$, the dynamics can be described as follows
(Figs.~\ref{subfig:ExRe-AdAVcPiQc40ICThreePhFuncOfg}-\ref{subfig:ExRe-SpSmVcPiQc40ICThreePhFuncOfg}):
First, a fast depletion of the initial three cavity photons into one photon, with timescale roughly determined by $\kappa_{c2}$. This can be seen by the sharp transition of the cavity excitation number from 3 to approximately 1 (red to blue in
Fig.~\ref{subfig:ExRe-AdAVcPiQc40ICThreePhFuncOfg}). Second, a slower depletion of the remaining cavity photon after a large number of Rabi exchanges between the qubit and the cavity, with timescale roughly determined by the decay rate of state $\ket{1_g,-}$. Essentially, since two-photon relaxation only allows pairs of exchange with the environment, the quantum state $\ket{1_g,-}$ acts like a dark state at $g=0$ (i.e. $\ket{1,g}$). As $g$ is increased, the decay rate of this state is barely modified up until $g/\nu_c \approx 0.5$ (See Fig.~\ref{subfig:PhenOpRabiDissVq8Em1Vc1Qc40FuncOfg}), consistent with the observed long-lived excitations in the qubit and cavity dynamics (Figs.~\ref{subfig:ExRe-AdAVcPiQc40ICThreePhFuncOfg}-\ref{subfig:ExRe-SpSmVcPiQc40ICThreePhFuncOfg}).

Steady state excitations have also been studied as a function of $g$ in Figs.~(\ref{subfig:ExRe-SSVcPiQc40ICTwoPhFuncOfg}-\ref{subfig:ExRe-SSVcPiQc40ICThreePhFuncOfg}). In the case of two initial photons, we observe that the steady state populations of the cavity and qubit increase non-monotonically with
increasing $g$, exhibiting a local maximum close to $g\approx\kappa_{c2}$. The case of three initial photons is more
complicated. For small $g<\kappa_{c2}$, one observes fast relaxation of two photons, while the remaining photon energy is transferred to the qubit at steady state. At intermediate values of $g$, the excitation is shared between the cavity and the qubit while at very large $g$, the qubit excitation saturates and the cavity photon population increases linearly
(Fig.~\ref{subfig:ExRe-SSVcPiQc40ICThreePhFuncOfg}). The overall increase observed in the steady state populations arises from the fact that the coupling in Eq.~(\ref{eqn:Model-Def of H_s,ef}) appears effectively as an incoherent drive on the cavity. Lastly, we note that the steady state quantities obtained from the Linbdlad formalism will become less accurate at large values of $g$, as one needs to account for the renormalization of the dissipator arising from the underlying system-bath formalism \cite{Beaudoin_Dissipation_2011}, resulting in a Bloch-Redfield master equation \cite{Redfield_Theory_1965}. Using Rayleigh-Schr\"odinger perturbation theory, however, one can show that disspator renormalizations are higher order in $g$ compared to the ones for the Hamiltonian. This leaves a window, at intermediate values of coupling, where the use of bare dissipators is still justified.


\textit{Acknowledgements}. We appreciate helpful discussions with Pengning Chao. This work was supported by the National Science Foundation under Grant No. DMR-1454836, Grant No. DMR 1420541, and Award EFMA-1640986. 
\clearpage
\appendix
\begin{center}
\textbf{Supplementary material: Quantum Rabi model with two-photon relaxation}
\end{center}
The structure of this supplemental material section is as follows. We first revisit a possible physical realization of two-photon relaxation for a cavity in Sec.~\ref{App:PhysReal}. The $Z_2$ symmetry of the Rabi model is discussed in detail in Sec.~\ref{App:ParSym}. In Sec.~\ref{App:ClRabiSpec}, we first revisit the spectrum of the closed Rabi model in its parity representation and extend the result to the open case up to phenomenological treatment of the opening. Section~\ref{App:EffHam} provides further discussion on the validity of the phenomenological treatment and its comparison with the full spectrum.

\section{Physical realization of two-photon relaxation}
\label{App:PhysReal}

In this section, we discuss potential physical realization of two-photon relaxation. In order to achieve a dissipator of the form $\mathcal{D}[\hat{a}^2]$ for the cavity, \textit{in principle}, one needs to engineer a cavity-bath coupling of the form
\begin{align}
\hat{H}_{\text{sb}}\approx\sum\limits_{k}\left[g_{2,k}^*\hat{a}^2\hat{b}_k^{\dag}+g_{2,k}(\hat{a}^\dag)^2\hat{b}_k\right],
\label{Eq:PhysReal-Hsb 1}
\end{align}  
where $g_{2,k}$ denotes the strength of the two-photon coupling between the cavity mode and mode $k$ of the bath. Such a coupling requires a three-wave mixing in which two cavity photons convert into one bath photon, but \textit{in practice}, it is realized by means of four-wave mixing including an additional pump tone. In what follows, we revisit the proposal by Wolinsky and Carmichael \cite{Wolinsky_Quantum_1988}, which have been recently experimentally implemented in the context of superconducting circuits \cite{Leghtas_Confining_2015}.

\begin{figure}[t!]
\centering \includegraphics[scale=0.40]{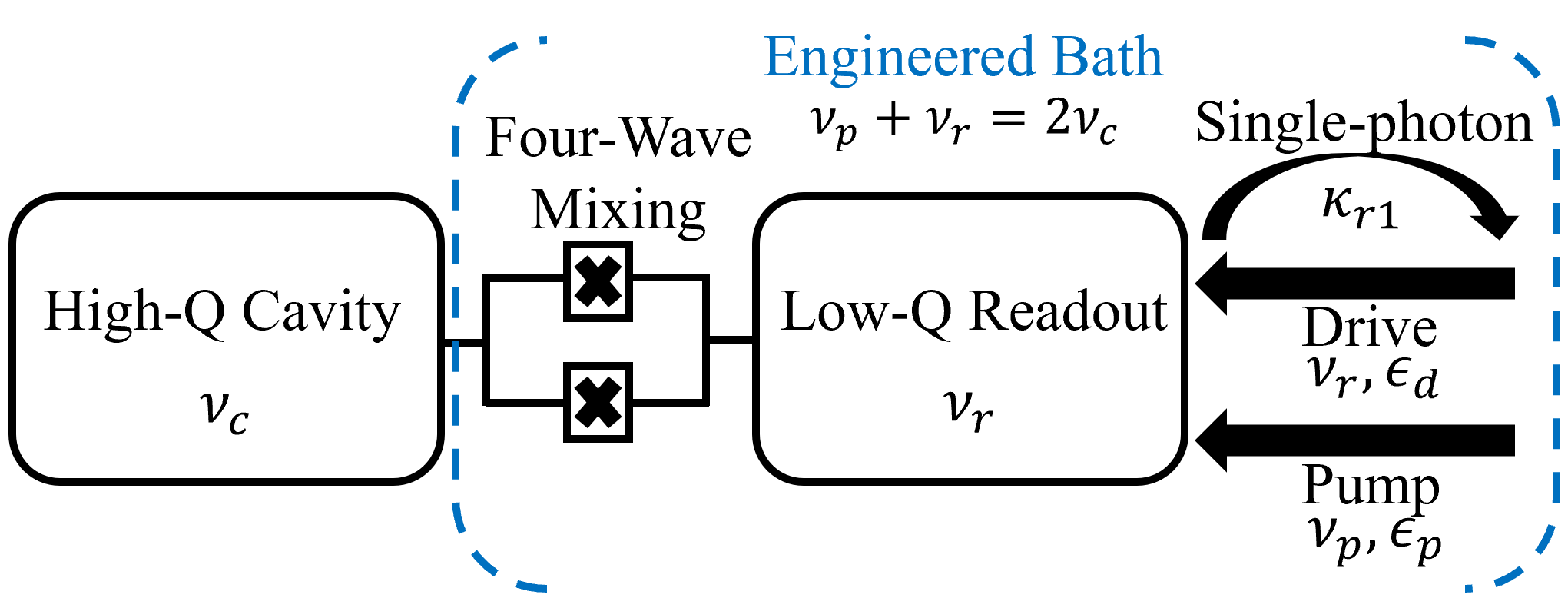}
\caption{Schematic of system consisting of a two-level system coupled to a single cavity mode with two-photon relaxation.}
\label{Fig:TwoPhotonRealization}
\end{figure}

In this scheme, the original cavity is required to have a very high quality factor in order to suppress the single-photon relaxation rate. The dominant two-photon relaxation is then achieved by engineering a quartic Kerr coupling to an additional linear cavity as shown in Fig.~\ref{Fig:TwoPhotonRealization}. In superconducting circuits, the quartic nonlinearity can be realized via a weakly nonlinear Josephson junction, in which the cosine potential can be well approximated by the lowest order quartic interaction. The auxiliary readout cavity, on the contrary, has to have very low quality factor in order to quickly dissipate the converted cavity photon pairs into the environment without significant back-scattering. 

To achieve the desired conversion, the frequencies of the pump and the cavities should satisfy the frequency matching condition 
\begin{align}
\nu_p+\nu_r=2\nu_c.
\label{Eq:PhysReal-FreqMatch}
\end{align}
Under condition~(\ref{Eq:PhysReal-FreqMatch}), the Hamiltonian for the overall system shown in Fig.~\ref{Fig:TwoPhotonRealization} can be approximated up to rotating-wave approximation as \cite{Leghtas_Confining_2015}
\begin{align}
\begin{split}
\hat{H}_{sb}&= g_2^*\hat{a}_c^2\hat{b}_r^{\dag}+g_2(\hat{a}_c^\dag)^2\hat{b}_r+\epsilon_d^*\hat{b}_r+\epsilon_d\hat{b}_r^{\dag}\\
&-\frac{\chi_{cc}}{2}(\hat{a}_c^{\dag})^2\hat{a}_c^2-\frac{\chi_{rr}}{2}(\hat{b}_r^{\dag})^2\hat{b}_r^2-\chi_{cr}\hat{a}_c^{\dag}\hat{a}_c\hat{b}_r^{\dag}\hat{b}_r,
\label{Eq:PhysReal-Hsb 2}
\end{split}
\end{align}
where $\hat{b}_r$ denotes the annihilation operator of the readout cavity. Importantly, in this scheme, we have achieved a pump-induced two-photon coupling as $g_2\equiv \chi_{rc}\xi_p/2$ with $\xi_p\equiv -i\epsilon_p/[(\nu_r-\nu_p)+i\kappa_{r1}]$ being the coherent amplitude of the readout cavity and $\kappa_{r1}$ the corresponding single-photon rate. The second line of Eq.~(\ref{Eq:PhysReal-Hsb 2}) keeps track of the additional self- and cross-Kerr interactions denoted by $\chi_{cc}$, $\chi_{rr}$ and $\chi_{rc}$ and can be considered as smaller perturbation compared to $g_2$. The additional drive tone at the readout frequency $\nu_r$ and amplitude $\epsilon_d$ is responsible to prepare the readout cavity with approximately one photon in order to fine-tune the desired conversion. Integrating out the readout degrees of freedom will result in an effective two-photon dissipator for the cavity as $\kappa_{c2}\mathcal{D}[\hat{a}_c^2]$ with relaxation rate $\kappa_{c2}$ given as \cite{Leghtas_Confining_2015}
\begin{align}
\kappa_{c2}\equiv \frac{\chi_{cr}}{\kappa_{r1}}|\xi_p|^2.
\end{align}

Finally, we note that in contrast to the aforementioned \textit{active} implementation that employs two drive tones, it might be also feasible to achieve a \textit{passive} implementation following similar ideas as the circuit-QED proposal by Felicetti et. al \cite{Felicetti_Two-Photon_2018} that is capable of achieving two-photon quantum Rabi model using a phase qubit coupled to a DC-SQUID.  
\section{$Z_2$ symmetry of the Rabi model}
\label{App:ParSym}
In this section, we revisit the $Z_2$ symmetry of the Rabi model and discuss the transformation that diagonalizes the Hamiltonian in its parity representation. 

We start with the Rabi Hamiltonian,
\begin{align}
\hat{H}_{\text{s}}\equiv \nu_c\hat{a}^{\dag}\hat{a}+\frac{\nu_q}{2}\hat{\sigma}^z+g\left(\hat{a}+\hat{a}^{\dag}\right)\left(\hat{\sigma}^-+\hat{\sigma}^+\right),
\label{Eq:ParSym-Rabi H}
\end{align}
where $\nu_c$ and $\nu_q$ are the cavity and qubit bare frequencies and $g$ represents the light--matter coupling strength. We define the overall parity operator $\hat{P}$ by its action on the spin and photonic degrees of freedom:
\begin{align}
\begin{cases}
\hat{P}^{\dag}\hat{\sigma}^-\hat{P}=-\hat{\sigma}^-\\
\hat{P}^{\dag}\hat{a}\hat{P}=-\hat{a}
\end{cases}.
\label{Eq:ParSym-Def of Par Op}
\end{align}
From~(\ref{Eq:ParSym-Def of Par Op}), one can derive the standard
properties of a parity operator, including involution
$\hat{P}^2=\hat{\mathbf{1}}$, unitarity
$\hat{P}^{\dag}\hat{P}=\hat{\mathbf{1}}$, and Hermiticity
$\hat{P}=\hat{P}^{\dag}$. The overall parity operator $\hat{P}$ can then be written as a product of the spin and photonic parity operators as
\begin{align}
\hat{P}=\hat{P}_q\hat{P}_c=e^{i\pi\hat{\sigma}^+\hat{\sigma}^-}e^{i\pi\hat{a}^{\dag}\hat{a}}=-\hat{\sigma}^{z}(-1)^{\hat{a}^{\dag}\hat{a}}.
\label{Eq:ParSym-Pi Sol}
\end{align}
The $Z_2$ symmetry of the Rabi Hamiltonian~(\ref{Eq:ParSym-Rabi H}) means that $\hat{H}_s$ remains invariant under the transformation
\begin{align}
\hat{P}^{\dag}\hat{H}_{s}\hat{P}=\hat{H}_s.
\label{Eq:ParSym-Def of Z2 Sym}
\end{align}

In the following, we rewrite the Rabi Hamiltonian such that the
$Z_2$ symmetry becomes explicit. In particular, we introduce
the following new set of bosonic creation and annihilation operators,
\begin{align}
\hat{b}\equiv\hat{\sigma}^x \hat{a},\ \hat{b}^{\dag}\equiv \hat{a}^{\dag}\hat{\sigma}^x,
\label{Eq:ParSym-Def b}
\end{align}
from which it follows that,
\begin{subequations}
\begin{align}
&[\hat{b},\hat{b}^{\dag}]=\hat{\sigma}^x[\hat{a},\hat{a}^{\dag}]\hat{\sigma}^x=1,
\label{Eq:ParSym-[b,b^dag]}\\
&\hat{b}^{\dag}\hat{b}=\hat{a}^{\dag}(\hat{\sigma}^x)^2\hat{a}=\hat{a}^{\dag}\hat{a}.
\label{Eq:ParSym-b^dag b}
\end{align}
\end{subequations}

Employing Eqs.~(\ref{Eq:ParSym-Pi Sol}) and~(\ref{Eq:ParSym-b^dag b}),
one can rewrite the Rabi Hamiltonian~(\ref{Eq:ParSym-Rabi H}) as
\cite{Casanova_Deep_2010}
\begin{align}
\hat{H}_{\text{s}}=\nu_c\hat{b}^{\dag}\hat{b}-\frac{\nu_q}{2}e^{i\pi\hat{b}^{\dag}\hat{b}}\hat{P}+g\left(\hat{b}+\hat{b}^{\dag}\right).
\label{Eq:ParSym-H_s transformed}
\end{align}
Since the parity and number operators commute, their eigenstates
provide a complete basis for the Hilbert space of the problem. Hence,
we define the number-parity basis $\ket{n,p}$, which satisfies:
\begin{subequations}
\begin{align}
&\hat{b}^{\dag}\hat{b}\ket{n,p}=n\ket{n,p}, \ n=0, 1, 2, \ldots\\
&\hat{P}\ket{n,p}=p\ket{n,p}, \ p=\pm 1.
\end{align} 
\end{subequations}
The advantage of working with the transformed Rabi
Hamiltonian~(\ref{Eq:ParSym-H_s transformed}) is that it is explicitly block-diagonal in the parity sector,
\begin{align}
\hat{H}_{\text{s}}=\begin{bmatrix}
\hat{H}_{\text{s},+} & 0\\
0 & \hat{H}_{\text{s},-}
\end{bmatrix},
\end{align}
with $\hat{H}_{\text{s},p}$ denoting the Hamiltonians of the even
($+$) and odd ($-$) parity subspaces, given by:
\begin{align}
\hat{H}_{\text{s},p} =\nu_c\hat{b}^{\dag}\hat{b}-\frac{p}{2}\nu_q e^{i\pi\hat{b}^{\dag}\hat{b}}+g(\hat{b}+\hat{b}^{\dag}), \quad p=\pm 1.
\label{Eq:ParSym-Def of H_(s,pm)} 
\end{align}

\section{Spectrum of the Rabi model}
\label{App:ClRabiSpec}

In this section, we study the spectrum of the Rabi
Hamiltonian~(\ref{Eq:ParSym-Def of H_(s,pm)}) and show that the
results can be extended to account for the open case up to phenomenological treatment. At last, we provide a comparison to the open JC, in which the counter-rotating terms in the coupling are dropped. We begin by considering the eigenvalue problem for each parity subspace,
\begin{align}
\hat{H}_{\text{s,p}}\ket{\Psi_{p}}=\omega_{p}\ket{\Psi_{p}},
\label{Eq:ClRabiSpec-EigProblem 1}
\end{align}
where $\omega_{p}$ and $\ket{\Psi_{p}}$ denote the eigenfrequency and eigenmode for each parity $p$, respectively.

Expanding the unknown wavefunctions $\ket{\Psi_{p}}$ in terms of the number-parity basis,
\begin{align}
\ket{\Psi_{p}}=\sum\limits_{m=0}^{\infty} c_{p,m}\ket{m,p},
\label{Eq:ClRabiSpec-Psi Ansatz}
\end{align}
and inserting this expansion into the eigenvalue
problem~(\ref{Eq:ClRabiSpec-EigProblem 1}), one obtains
\begin{align}
\begin{split}
&\sum\limits_{m=0}^{\infty}\left[m\nu_c-\frac{p}{2}(-1)^{m}\nu_q-\omega_{p}\right]c_{p,m}\ket{m,p}\\
+&\sum\limits_{m=0}^{\infty}gc_{p,m}\left(\sqrt{m}\ket{m-1,p}+\sqrt{m+1}\ket{m+1,p}\right)=0.
\end{split}
\label{Eq:ClRabiSpec-EigProblem 2}
\end{align}
Reindexing the sums and employing the linear independence of the number-parity basis $\{\ket{m,p}\}$, we obtain the recurrence relation for $m\geq 1$ as
\begin{subequations}
\begin{align}
\begin{split}
\alpha_{p,m}c_{p,m}&+\beta_{p,m}c_{p,m+1}+\gamma_{p,m}c_{p,m-1}=0,
\end{split}
\label{Eq:ClRabiSpec-RecRel n>=1}
\end{align}
with $\alpha_{p,m}$, $\beta_{p,m}$ and $\gamma_{p,m}$ defined as
\begin{align}
&\alpha_{p,m}\equiv \omega_{p}-m\nu_c+\frac{p}{2}(-1)^m\nu_q,
\label{Eq:ClRabiSpec-Def of alpha_p,n}\\
&\beta_{p,m}\equiv-\sqrt{m+1}g,
\label{Eq:ClRabiSpec-Def of beta_p,n}\\
&\gamma_{p,m}\equiv-\sqrt{m}g,
\label{Eq:ClRabiSpec-Def of gamma_p,n}
\end{align}
and with initial conditions,
\begin{align}
\alpha_{p,0}c_{p,0}+\beta_{p,0}c_{p,1}=0.
\label{Eq:ClRabiSpec-RecRel n=1}
\end{align}
\end{subequations}
The recursion relation~(\ref{Eq:ClRabiSpec-RecRel n>=1}) and
the associated initial condition~(\ref{Eq:ClRabiSpec-RecRel n=1}) uniquely determine $c_{p,m}$ in terms of $c_{p,0}$ for arbitrary $m$.

Next, we need to obtain an equation to determine the eigenfrequencies
$\omega_p$. Note that the recursion relation above can be represented in matrix form, $\mathbf{M}_p\mathbf{c}_p=0$, in terms of the infinite-dimensional tridiagonal matrix:
\begin{align}
\mathbf{M}_{p}\equiv
\begin{bmatrix}
\alpha_{p,0} & \beta_{p,0} & 0 & 0 & \ldots \\
\gamma_{p,1} & \alpha_{p,1} & \beta_{p,1} & 0 & \ldots \\
0 & \gamma_{p,2} & \alpha_{p,2} & \beta_{p,2} & \ldots \\
\vdots & \vdots & \vdots & \ddots & \ldots 
\end{bmatrix}.
\label{Eq:ClRabiSpec-Def of Mp}
\end{align}
Hence, the roots of the determinant of $\mathbf{M}_p$ yield the
eigenfrequencies corresponding to subspace $p$. Denoting the
determinant $G_{p,m}\equiv \det (\mathbf{M}_{p,m})$, where
$\mathbf{M}_{p,m}$ is an $(m+1)\times (m+1)$ truncation of
$\mathbf{M}_p$, and employing a Laplace expansion, one can obtain a
recursive relation for the determinant,
\begin{align}
G_{p,m}=\alpha_{p,m} G_{p,m-1}-\beta_{p,m-1}\gamma_{p,m}G_{p,m-2},
\label{Eq:ClRabiSpec-RecRelG n>=1}
\end{align}
with initial condition $G_{p,0}=\alpha_{p,0}$ and
$G_{p,1}=\alpha_{p,0}
\alpha_{p,1}-\beta_{p,0}\gamma_{p,1}$. Defining,
\begin{align}
G_p\equiv\lim\limits_{m\to\infty}G_{p,m},
\label{Eq:ClRabiSpec-Def of Gp}
\end{align}
it follows that the $n$th eigenfrequency in each subspace is given by the $n$th root of $G_p$, obtained by solving $G_p(\omega_{np})=0$. The corresponding eigenmode is found by replacing $\omega_{np}$ in Eq.~(\ref{Eq:ClRabiSpec-Def of alpha_p,n}) to find $\alpha_{np,m}$ and related quantities accordingly.
\begin{figure}[t!]
\centering
\subfloat[\label{subfig:PhenOpSpecRabiVsJCVq8Em1Vc1Qc40FuncOfg}]{%
\includegraphics[scale=0.32]{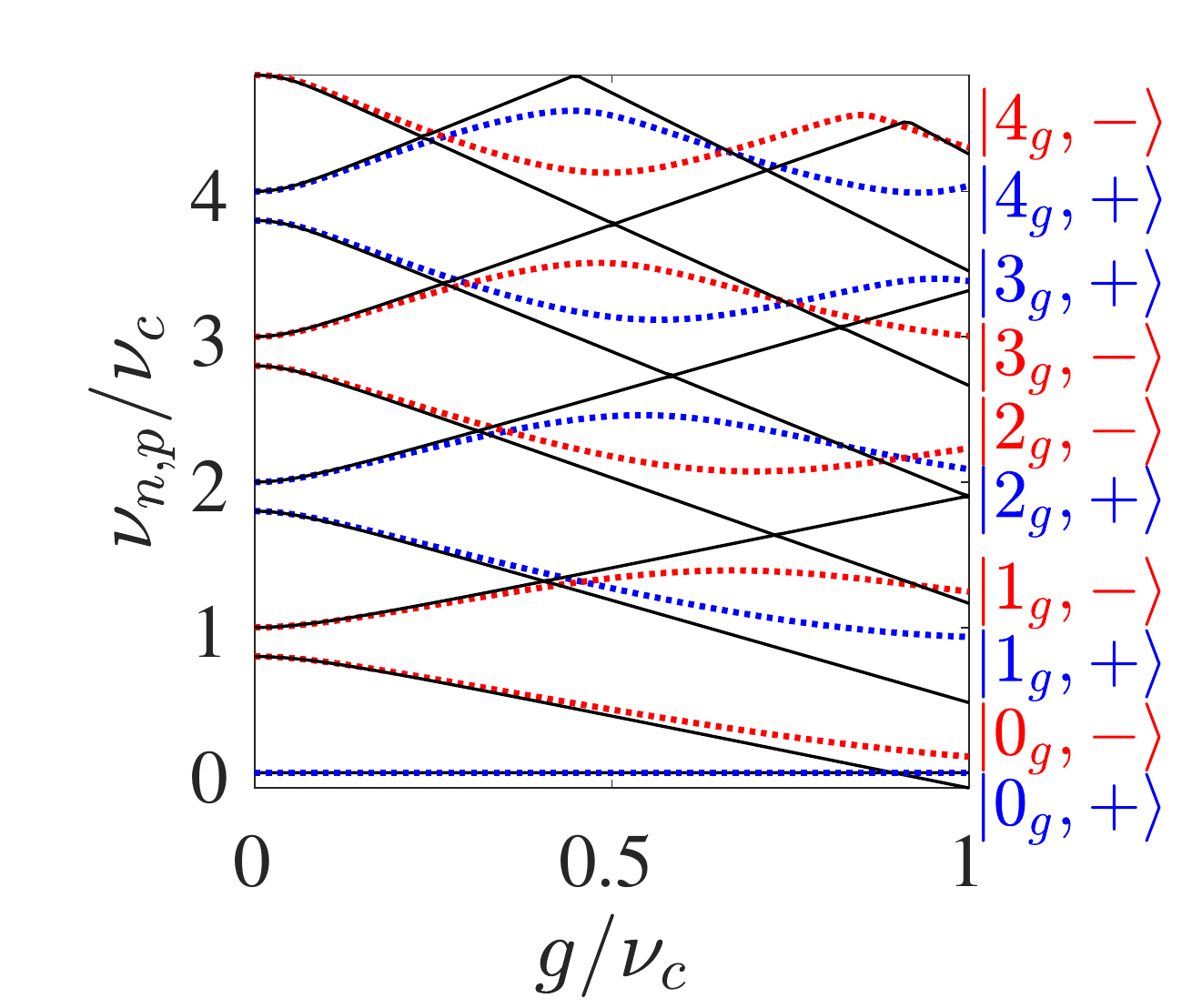}%
}
\subfloat[\label{subfig:PhenOpDissRabiVsJCVq8Em1Vc1Qc40FuncOfg}]{%
\includegraphics[scale=0.32]{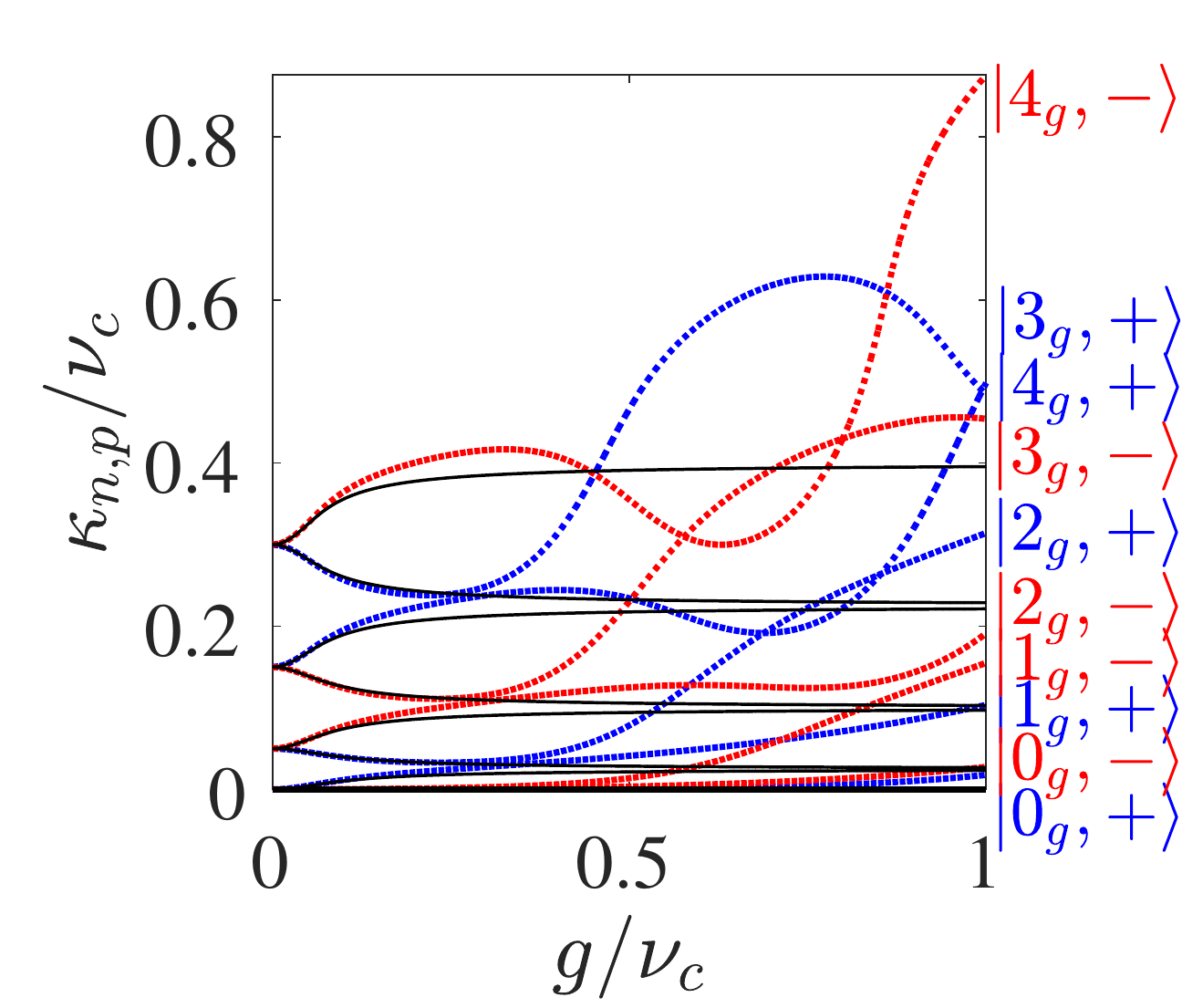}%
}
\caption{Phenomenological open eigenfrequencies
  $\omega_{n,p}\equiv \nu_{n,p}-i\kappa_{n,p}$ for $\nu_q=0.8\nu_c$ and $\kappa_{c2}=\nu_c/40$.  a) Real part (frequency) of the spectrum and b) imaginary part (decay rate) of the spectrum as a function of light-matter coupling $g$. The red/blue dotted lines show the result obtained from the Rabi model~(\ref{Eq:ClRabiSpec-Def of H_s,ef}), while the solid black lines show the JC result obained from Eq.~(\ref{Eq:ClRabiSpec-H_JC,ef}). The labels $\ket{n_g,\pm}$ in a) and b) are ordered based on values at $g=0$. The frequencies in a) are plotted relative to the ground state $\ket{0_g,+}$.}
\label{Fig:PhenOpSpecRabiVsJC} 
\end{figure}

If the relaxation is accounted phenomenologically, we employ the effective Hamilatonian as
\begin{align}
\hat{H}_{\text{s,ef}}=\nu_c\hat{b}^{\dag}\hat{b}-\frac{\nu_q}{2}e^{i\pi\hat{b}^{\dag}\hat{b}}\hat{P}+g\left(\hat{b}+\hat{b}^{\dag}\right)-i\kappa_{c2} (\hat{b}^{\dag})^2\hat{b}^2.
\label{Eq:ClRabiSpec-Def of H_s,ef}
\end{align}
Note that the additional term $-i\kappa_{c2}(\hat{b}^{\dag})^2\hat{b}^2$ is diagonal in the number basis with matrix elements $-i m(m-1)\kappa_{c2}$. Therefore, following the same steps as in Eqs.~(\ref{Eq:ClRabiSpec-Psi Ansatz}-\ref{Eq:ClRabiSpec-EigProblem 2}), we find similar recursion
relations as Eqs.~(\ref{Eq:ClRabiSpec-RecRel n>=1}) and~(\ref{Eq:ClRabiSpec-RecRelG n>=1}), for the eigenmodes and spectrum respectively, where  the diagonal coefficient $\alpha_{p,m}$ gets replaced due to the dissipative contribution as
\begin{align}
\begin{split}
\alpha_{p,m}&\rightarrow \alpha_{p,m}+im\left(m-1\right)\kappa_{c2}\\
&=\omega_{p}-m\nu_c+\frac{p}{2}(-1)^m\nu_q+im\left(m-1\right)\kappa_{c2}
\end{split},
\label{Eq:ClRabiSpec-Def of alpha_p,n^ph}
\end{align}
while the off-diagonoal coefficients $\beta_{p,m}$ and $\gamma_{p,m}$ remain intact. As expected, the eigenfrequencies $\omega_{np}$ become complex.  An example comparing the closed and open spectrum is shown in
Fig.~2 of the paper.

Lastly, we study the Jaynes-Cummings (JC) model with two-photon relaxation and employ the \textit{analytical} solutions to its spectrum to compare to and benchmark the result obtained from our recursion relations. Consider the phenomenological effective JC Hamiltonian with two-photon relaxation [analogous to Eq.~(\ref{Eq:ClRabiSpec-Def of H_s,ef})] as  
\begin{align}
\hat{H}_{\text{JC,ef}}=\nu_c\hat{a}^{\dag}\hat{a}+\frac{\nu_q}{2}\hat{\sigma}^z+g(\hat{a}\hat{\sigma}^{+}+\hat{a}^{\dag}\hat{\sigma}^-)-i\kappa_{c2}(\hat{a}^\dag)^2\hat{a}^2.
\label{Eq:ClRabiSpec-H_JC,ef}
\end{align}
where the counter-rotating terms are dropped. Hamiltonian~(\ref{Eq:ClRabiSpec-H_JC,ef}) commutes with the total excitation number $\hat{N}\equiv \hat{a}^{\dag}\hat{a}+\hat{\sigma}^+\hat{\sigma}^-$. Therefore, it is block-diagonal consisting of the singlet $\ket{0,g}$ with eigenfrequency $\omega_{0,+}=-\nu_q/2$ and doublets $\{\ket{n-1,e},\ket{n,g}\}$ as 
\begin{widetext}
\begin{align}
\begin{bmatrix}
(n-1)\nu_c+\nu_q/2-i(n-1)(n-2)\kappa_{c2} & g\sqrt{n}\\
g\sqrt{n} & n\nu_c-\nu_q/2-i\kappa_{c2}(n-1)n
\end{bmatrix}.
\label{Eq:ClRabiSpec-H_JC 2X2}
\end{align}
\end{widetext}
The eigenfrequencies of Eq.~(\ref{Eq:ClRabiSpec-H_JC 2X2}) can be immediately found as
\begin{subequations}
\begin{align}
\begin{split}
\omega^{\text{JC}}_{n,(-1)^n}&=\left(n-\frac{1}{2}\right)\nu_c-i(n-1)^2\kappa_{c2}\\
+&\frac{1}{2}\sqrt{[\nu_c-\nu_q-2i(n-1)\kappa_{c2}]^2+4g^2n},
\end{split}
\label{Eq:ClRabiSpec-JC Om_(n)}
\end{align}
\begin{align}
\begin{split}
\omega^{\text{JC}}_{n-1,(-1)^n}&=\left(n-\frac{1}{2}\right)\nu_c-i(n-1)^2\kappa_{c2}\\
-&\frac{1}{2}\sqrt{[\nu_c-\nu_q-2i(n-1)\kappa_{c2}]^2+4g^2n}.
\end{split}
\label{Eq:ClRabiSpec-JC Om_(n-1)}
\end{align}
\end{subequations}
where the labels are chosen to be consistent with our convention for the Rabi eigenvectors $\omega_{np}$. Figure~\ref{Fig:PhenOpSpecRabiVsJC} provides a comparison between the phenomenological spectrum obtained from the Rabi and JC models. Besides the well-known spurious level crossings that is observed in the real part of the JC spectrum (Fig.~\ref{subfig:PhenOpSpecRabiVsJCVq8Em1Vc1Qc40FuncOfg}), we observe that the JC model generate decay rates that \textit{plateau} as $g$ is enhanced and hence completely mischaracterizes the interplay of the coupling and the two-photon relaxation at ultrastrong coupling (Fig.~\ref{subfig:PhenOpDissRabiVsJCVq8Em1Vc1Qc40FuncOfg}).

\section{Full effective Hamiltonian}
\label{App:EffHam}

In the previous section, we obtained a rather simple generalization of the recursion relation for the spectrum to the open case up to phenomenological treatment of the opening [Eq.~(\ref{Eq:ClRabiSpec-Def of alpha_p,n^ph})]. We have employed the resulting approximate result to analyze the excitation-relaxation dynamics discussed in the main paper. The main purpose of this section is to provide more discussion on the validity of such an approximation. In subsection~\ref{SubApp:GenDis}, we first revisit the possibility of extending the effective Hamiltonian approach such that it accounts for the collapse terms as well. The full effective Hamiltonian belongs to a larger Hilbert space, hence an immediate question is the possibility of decomposition of the full spectrum. This is discussed in subsection~\ref{SubApp:SpecDecom}. Subsection~\ref{SubApp:DissTLS} applies the full effective Hamiltonian approach and the spectral decomposition on the simplest case of a two-level system with relaxation, and provides important insight on the role of collapse term in renormalizaiton of eigenmodes and spectrum. Finally, in Sec.~\ref{SubApp:EfHamRabi}, we provide a comparison between the spectrum derived from phenomenological and full effective Hamiltonian approaches for our system.        

\subsection{General discussion}
\label{SubApp:GenDis}

Here, we revisit the derivation of the full effective Hamiltonian approach for open quantum systems that was first introduced by Yi et. al \cite{Yi_Effective_2001}. Consider a generic Lindblad equation   
\begin{subequations}
\begin{align}
\dot{\hat{\rho}}_s=-i[\hat{H}_s,\hat{\rho}_s]+\sum\limits_{\lambda}2\gamma_{\lambda}\mathcal{D}[\hat{C}_{s,\lambda}]\hat{\rho}_s,
\label{Eq:EffHam-Lindblad 1}
\end{align}
where $\hat{\rho}_s$ and $\hat{H}_s$ are the system density matrix and Hamiltonian, respectively. Moreover, we assume a set of dissipation channels, with relaxation rate $\gamma_{\lambda}$ and collapse operator $\hat{C}_{\lambda}$ of the form
\begin{align}
\mathcal{D}[\hat{C}_{s,\lambda}]\equiv \hat{C}_{s,\lambda}(\bullet)\hat{C}_{s,\lambda}^{\dag}-\frac{1}{2}\left\{\hat{C}_{s,\lambda}^{\dag}\hat{C}_{s,\lambda},(\bullet)\right\}.
\label{Eq:EffHam-Def of D[c]}
\end{align}
\end{subequations}

The diagonal terms (anti-commutator) in Eq.~(\ref{Eq:EffHam-Def of D[c]}) can be expressed as effective decay if we define an effective system Hamiltonian as
\begin{subequations}
\begin{align}
\hat{H}_{\text{s,ef}}\equiv \hat{H}_s-\sum\limits_{\lambda}i\gamma_{\lambda}\hat{C}_{s,\lambda}^{\dag}\hat{C}_{s,\lambda},
\label{Eq:EffHam-Def of H_s,ef}
\end{align}
in terms of which the original Lindblad Eq.~(\ref{Eq:EffHam-Lindblad 1}) can be recast into
\begin{align}
\dot{\hat{\rho}}_s=-i\left[\hat{H}_{\text{s,ef}}\hat{\rho}_s-\hat{\rho}_s(t)\hat{H}_{\text{s,ef}}^{\dag}\right]+\sum\limits_{\lambda}2\gamma_{\lambda}\hat{C}_{s,\lambda}\hat{\rho}_s\hat{C}_{s,\lambda}^{\dag}.
\label{Eq:EffHam-Lindblad 2}
\end{align}
\end{subequations}
  
The idea of the full effective Hamiltonian approach is to map the Lindblad Eq.~(\ref{Eq:EffHam-Lindblad 2}) to an effective Schrodinger equation. This is achieved by introducing an auxiliary system with the same Hilbert space size as the original system, which plays the role of adjoint (left) quantum states. This allows extending the size of the composite Hilbert space to that of the original Lindbladian. We therefore introduce a full orthonormal basis $\left\{\ket{n_s}\ket{n_a}\right\}$ and define the following full wavefunction
\begin{align}
\ket{\Psi_{\hat{\rho}}(t)}\equiv \sum\limits_{n_s,n_a}\rho_{s,n_s n_a}(t)\ket{n_s}\ket{n_a},
\label{Eq:EffHam-Def of Psi_rho}
\end{align}
where $\rho_{s,n_s n_a}(t)$ are the matrix elements of the system density matrix defined as $\rho_{s,n_s n_a}(t)\equiv\bra{n_s}\hat{\rho}_{s}(t)\ket{n_a}$. We then seek an effective Hamiltonian such that the Schr\"odinger dynamics of the pure wavefunction $\ket{\Psi_{\hat{\rho}_s}(t)}$ matches the original Lindblad dynamics in terms of the matrix elements $\rho_{s,n_sn_a}(t)$. 

The resulting full effective Hamiltonian for the system can be found as \cite{Yi_Effective_2001}
\begin{subequations}
\begin{align}
\hat{H}_{\text{u,ef}}\equiv\hat{H}_{\text{s,ef}}-\hat{H}_{\text{a,ef}}+i\sum\limits_{\lambda}2\gamma_{\lambda}\hat{C}_{s,\lambda}\hat{C}_{a,\lambda},
\label{Eq:EffHam-Def of H_u}
\end{align}
where $\hat{H}_{\text{a,ef}}$ and $\hat{C}_{a,\lambda}$ are the effective Hamiltonian and collapse operator for the auxiliary system that satisfy
\begin{align}
&\bra{n_a'}\hat{H}_{\text{a,ef}}\ket{n_a}=\bra{n_s}\hat{H}_{\text{s,ef}}^{\dag}\ket{n_s'},
\label{Eq:EffHam-Def of H_a,ef}\\
&\bra{n_a'}\hat{C}_{a,\lambda}\ket{n_a}=\bra{n_s}\hat{C}_{s,\lambda}^{\dag}\ket{n_s'}.
\label{Eq:EffHam-Def of C_a}
\end{align}
\end{subequations}
Consequently, in this approach, we seek a solution to the wavefunction as
\begin{align}
\ket{\Psi_{\hat{\rho}}(t)}=e^{-i\int_{0}^{t}\hat{H}_{\text{u,ef}}dt'}\ket{\Psi_{\hat{\rho}}(0)},
\label{Eq:EffHam-Sol of |Psi_rho>}
\end{align}
from which one can infer the solution for $\hat{\rho}_s(t)$.

Equation~(\ref{Eq:EffHam-Def of H_u}) clarifies the distinction between the phenomenological effective Hamiltonian~(\ref{Eq:EffHam-Def of H_s,ef}) and the full effective Hamiltonian more clear. In the phenomenological treatment, the system spectrum is determined by the eigenvalues of $\hat{H}_{\text{s,ef}}$, where the additional coupling caused by the collapse operators between the left and right states are neglected. 

\subsection{Spectral decomposition}
\label{SubApp:SpecDecom}

At first sight, it seems that the additional coupling (last term) in the full effective Hamitlonian~(\ref{Eq:EffHam-Def of H_u}) breaks the possibility of spectral decomposition, i.e. writing the full spectrum as linear sum of each constituent. It has been shown that for quadratic Lindbladians it is feasible to obtain a reduced effective spectrum for the system in terms of the full spectrum \cite{Prosen_Spectral_2010} due to the following self-conjugation symmetry. Note that under the transformation 
\begin{subequations}
\begin{align}
\hat{C}_{s,\lambda}\leftrightarrow \hat{C}_{a,\lambda},
\label{Eq:EffHam-C_s<->C_a}
\end{align}
which maps the original system to the adjoint system, we obtain from Eq.~(\ref{Eq:EffHam-Def of H_u}) the self-conjugation symmetry
\begin{align}
\hat{H}_{\text{u,ef}}\rightarrow - \hat{H}_{\text{u,ef}}^*.
\label{Eq:EffHam-Sym of H_u,ef}
\end{align}
\end{subequations}
The consequence of symmetry~(\ref{Eq:EffHam-Sym of H_u,ef}) for a quadratic Lindbladian is that the full eigenfrequencies can always be effectively decomposed as 
\begin{align}
\left\{\omega_{u,mn}\right\}=\left\{\omega_{m}-\omega_{n}^*\right\}, 
\label{Eq:EffHam-Om_u,mn}
\end{align}
where $\{\omega_{u,mn}\}$ is the set of full eigenfrequencies and $\{\omega_n\}$ is the set of reduced eigenfrequencies for the system that accounts for the collapse terms by construct. Assuming that we have access to the full spectrum, analytically or numerically, we face the problem of obtaining $N$ reduced complex eigenfrequencies for the system in terms of the corresponding $N^2$ full complex eigenfrequencies, with $N$ being the cut-off number for the Hilbert space.  

We note that the elementwise Eq.~(\ref{Eq:EffHam-Om_u,mn}) can be compactly represented as a Tensor Rank Decomposition \cite{Kolda_Tensor_2009} problem of the form 
\begin{subequations}
\begin{align}
&\mathbf{\Omega}_u=\mathbf{\Omega}\otimes\mathbf{1}_a-\mathbf{1}_s\otimes\mathbf{\Omega}^*,
\label{Eq:EffHam-TRD for Om}
\end{align}
or in terms of real frequency and decay rate tensors as
\begin{align}
&\mathbf{V}_u=\mathbf{V}\otimes\mathbf{1}_a-\mathbf{1}_s\otimes\mathbf{V},
\label{Eq:EffHam-TRD for V}\\
&\mathbf{K}_u=\mathbf{K}\otimes\mathbf{1}_a+\mathbf{1}_s\otimes\mathbf{K},
\label{Eq:EffHam-TRD for K}
\end{align}
\end{subequations}
with $\mathbf{1}$, $\mathbf{V}$, $\mathbf{K}$ being the identity, frequency and decay matrices for each sector. The solution for $\mathbf{V}$ in terms of $\mathbf{V}_u$ is unique up to a overall translation, e.g. if $\mathbf{V}$ is a solution then $\mathbf{V}+c\mathbf{1}$ is also a solution for $c\in \mathbb{R}$. This is not an issue, since the overall constant only changes the global reference point with respect to which the frequency levels in $\mathbf{V}$ are evaluated, while leaves the $\textit{physical}$ difference that appears in the time-evolution of the system invariant.

In contrast to the spectrum, it is not in general possible to tensor decompose the full eigenmodes into a reduced set of eigenmodes for the system. As an example, a full quantum state 
\begin{align}
\ket{\Psi_{u,mn}}=\ket{m_s}\ket{n_a}\pm \ket{n_s}\ket{m_a},
\label{Eq:RabiSpec-No TRD for Psi_u,mn}
\end{align}  
is a valid possibility for the full eigenmode since it respects the symmetry~(\ref{Eq:EffHam-Sym of H_u,ef}) of the full Hamiltonian. However, a full quantum state like Eq.~(\ref{Eq:RabiSpec-No TRD for Psi_u,mn}) can not be expresed as a tensor product of two independent states in each sector, i.e. $\ket{\Psi_{u,mn}}\neq \ket{\Psi_{s,m}}\ket{\Psi_{a,n}}$.

\subsection{A two-level system with relaxation}
\label{SubApp:DissTLS}

Here, we study the case of a two-level system with relaxation. The aim of this subsection is to apply the tools from Secs.~\ref{SubApp:GenDis} and~\ref{SubApp:SpecDecom} on a rather simple example that can be handled analytically. The main outcome of this calculation is that the phenomenological treatment of the opening captures the spectrum correctly, but not the ground state and the modal structure in general. Hence, using the phenomenological treatment provides a good approximation of the spectrum (in this case exact), but cannot be used to draw conclusions about the eigenmodes and in particular the steady state.         

Consider the Lindblad equation
\begin{align}
\dot{\hat{\rho}}(t)=-i[\hat{H}_q,\hat{\rho}(t)]+2\gamma_q\mathcal{D}[\hat{\sigma}^-]\hat{\rho}(t),
\label{Eq:DissTLS-Lindblad Eq}
\end{align}
where $\hat{H}_q$ and $\mathcal{D}[\hat{\sigma}^-]$ are the Hamiltonian and the dissipator superoperator given in terms of spin-$1/2$ Pauli matrices as
\begin{align}
&\hat{H}_q\equiv \nu_q\hat{\sigma}^+\hat{\sigma}^-,
\label{Eq:DissTLS-Def of Hq}\\
&\mathcal{D}[\hat{\sigma}^-](\bullet)\equiv \hat{\sigma}^-(\bullet)\hat{\sigma}^+-\frac{1}{2}\left\{\hat{\sigma}^+\hat{\sigma}^-,(\bullet)\right\}.
\label{Eq:DissTLS-Def of D[sig-]}
\end{align}
In terms of the phenomenological effective Hamiltonian
\begin{align}
&\hat{H}_{q,\text{ef}}\equiv\hat{H}_q-i\gamma_q\hat{\sigma}^+\hat{\sigma}^-=\omega_q\hat{\sigma}^+\hat{\sigma}^-,
\label{Eq:DissTLS-Def of H_ef}
\end{align}
with complex frequency $\omega_q\equiv \nu_q-i\gamma_q$, we can reexpress Eq.~(\ref{Eq:DissTLS-Lindblad Eq}) as
\begin{align}
\dot{\hat{\rho}}(t)=-i\left[\hat{H}_{q,\text{ef}}\hat{\rho}(t)-\hat{\rho}(t)\hat{H}_{q,\text{ef}}^{\dag}\right]+2\gamma_q\hat{\sigma}^-\hat{\rho}(t)\hat{\sigma}^+.
\label{Eq:DissTLS-Lindblad Eq 2}
\end{align}

Next, we discuss the solution to the Lindblad Eq.~(\ref{Eq:DissTLS-Lindblad Eq 2}) using Effective Hamiltonian approach. The full effective Hamiltonian includes an auxiliary TLS and reads
\begin{align}
\hat{H}_{\text{u,ef}}=(\nu_q-i\gamma_q)\hat{\sigma}^{+}\hat{\sigma}^--(\nu_q+i\gamma_q)\hat{\Sigma}^{+}\hat{\Sigma}^-+2i\gamma_q \hat{\sigma}^{-}\hat{\Sigma}^{-},
\label{Eq:DissTLS-H_u,ef}
\end{align}
where we have denoted the spin operators of the auxiliary system with $\hat{\Sigma}$. In the following we prepare the system in a generic pure state 
\begin{align}
\begin{split}
\ket{\Psi_{\hat{\rho}}(0)}&=\rho_{ee}(0)\ket{e_s}\ket{e_a}+\rho_{eg}(0)\ket{e_s}\ket{g_a}\\
&+\rho_{ge}(0)\ket{g_s}\ket{e_a}+\rho_{gg}(0)\ket{g_s}\ket{g_a},
\end{split}
\label{Eq:DissTLS-IC for Psi(0)}
\end{align}
and solve for the Schrodinger equation
\begin{align}
i\partial_t \ket{\Psi_{\hat{\rho}}(t)}=\hat{H}_{\text{u,ef}}\ket{\Psi_{\hat{\rho}}(t)}.
\label{Eq:DissTLS-Schro Eq}
\end{align}
Expressing the full effective Hamiltonian~(\ref{Eq:DissTLS-H_u,ef}) in the basis $\{\ket{s_{\sigma}}\ket{s_{\Sigma}}\}$ we find
\begin{align}
\mathbf{H}_{\text{u,ef}}\equiv
\begin{bmatrix}
-2i\gamma_q & 0 & 0 & 0\\
0 & \nu_q-i\gamma_q & 0 & 0\\
0 & 0 & -\nu_q-i\gamma_q & 0\\
2i\gamma_q & 0 & 0 & 0
\end{bmatrix}.
\label{Eq:DissTLS-Rep of H_u,ef}
\end{align}
We can then directly compute the matrix representation of time-evolution operator $\hat{U}(t)\equiv e^{-i\hat{H}_{\text{u,ef}} t}$ as
\begin{align}
\mathbf{U}(t)=
\begin{bmatrix}
e^{-2\gamma_q t} & 0 & 0 & 0\\
0 & e^{-\gamma_q t-i\nu_q t} & 0 & 0\\
0 & 0 & e^{-\gamma_q t+i\nu_q t} & 0\\
1-e^{-2\gamma_q t} & 0 & 0 & 1 
\end{bmatrix}.
\label{Eq:DissTLS-Rep of U(t)}
\end{align}
The solution for $\ket{\Psi_{\hat{\rho}}(t)}$ is obtained as
\begin{align}
\ket{\Psi_{\hat{\rho}}(t)}=\hat{U}(t)\ket{\Psi_{\hat{\rho}}(0)},
\end{align}
which can be expressed element-wise as\begin{subequations}
\begin{align}
&\rho_{ee}(t)=\rho_{ee}(0)e^{-2\gamma_q t},
\label{Eq:DissTLS-rho_ee Sol}\\
&\rho_{eg}(t)=\rho_{eg}(0)e^{-\gamma_q t-i\nu_q t},
\label{Eq:DissTLS-rho_eg Sol}\\
&\rho_{ge}(t)=\rho_{ge}(0)e^{-\gamma_q t+i\nu_q t}.
\label{Eq:DissTLS-rho_ge Sol}\\
&\rho_{gg}(t)=\rho_{ee}(0)[1-e^{-2\gamma_q t}]+\rho_{gg}(0).
\label{Eq:DissTLS-rho_gg Sol}
\end{align}
\end{subequations}

It is instructive to clarify the role of collapse term in this rather simple example. The full effective Hamiltonian~(\ref{Eq:DissTLS-Rep of H_u,ef}) is lower triangular, i.e. that the collapse term does not change the eigenfrequencies, but rather the eigenvectors of the system. Regardless of the inclusion of the collapse terms, the eigenvalues of $\hat{H}_{\text{u,ef}}$ read
\begin{subequations}
\begin{align}
&\omega_{u,ee}=-2i\gamma_q, 
\label{Eq:DissTLS-Om_u,ee}\\
&\omega_{u,eg}=\nu_q-i\gamma_q,
\label{Eq:DissTLS-Om_u,eg}\\
&\omega_{u,ge}=-\nu_q-i\gamma_q,
\label{Eq:DissTLS-Om_u,ge}\\
&\omega_{u,gg}=0.
\label{Eq:DissTLS-Om_u,gg}
\end{align}
\end{subequations}
On the other hand, if the collapse term is neglected, the eigenvectors are simply the starting excitation basis $\{\ket{e_s}\ket{e_a}, \ket{e_s}\ket{g_a}, \ket{g_s}\ket{e_a}, \ket{g_s}\ket{g_a}\}$; while the eigenvectors of Eq.~(\ref{Eq:DissTLS-Rep of H_u,ef}) read
\begin{subequations}
\begin{align}
&\ket{\Psi_{u,ee}}=\ket{e_s}\ket{e_a},
\label{Eq:DissTLS-Psi_u,gg}\\
&\ket{\Psi_{u,eg}}=\ket{e_s}\ket{g_a},
\label{Eq:DissTLS-Psi_u,gg}\\
&\ket{\Psi_{u,ge}}=\ket{g_s}\ket{g_a},
\label{Eq:DissTLS-Psi_u,gg}\\
&\ket{\Psi_{u,gg}}=\frac{1}{\sqrt{2}}\left(\ket{g_s}\ket{g_a}-\ket{e_s}\ket{e_a}\right).
\label{Eq:DissTLS-Psi_u,gg}
\end{align}
\end{subequations}
Therefore, the last eigenvector changes from $\ket{g_s}\ket{g_a}$ to $1/\sqrt{2}(\ket{g_s}\ket{g_a}-\ket{e_s}\ket{e_a})$. The modification of the eigenvector has a significant impact on the dynamics and the steady state in particular. For example, without the collapse term, the solution for $\rho_{gg}(t)$ in Eq.~(\ref{Eq:DissTLS-rho_gg Sol}) is wrongly reduced to $\rho_{gg}(t)=\rho_{gg}(0)$, which breaks the conservation of probability.

At last, we discuss the application of tensor rank decomposition of the full spectrum~(\ref{Eq:DissTLS-Om_u,ee}-\ref{Eq:DissTLS-Om_u,gg}) into a reduced spectrum as discussed in Sec.~(\ref{App:EffHam}). Defining the reduced spectrum 
\begin{align}
\omega_g=0, \quad \omega_e\equiv \nu_q-i\gamma_q,
\end{align}
it is clear that the full spectrum is simply found as $\omega_{u,mn}=\omega_{m}-\omega_{n}^*$ for $m,n \in \{e,g\}$.

\subsection{Rabi model with two-photon relaxation}
\label{SubApp:EfHamRabi}
\begin{figure}[t!]
\centering
\subfloat[\label{subfig:QuOpRabiSpecPP}]{%
\includegraphics[scale=0.40]{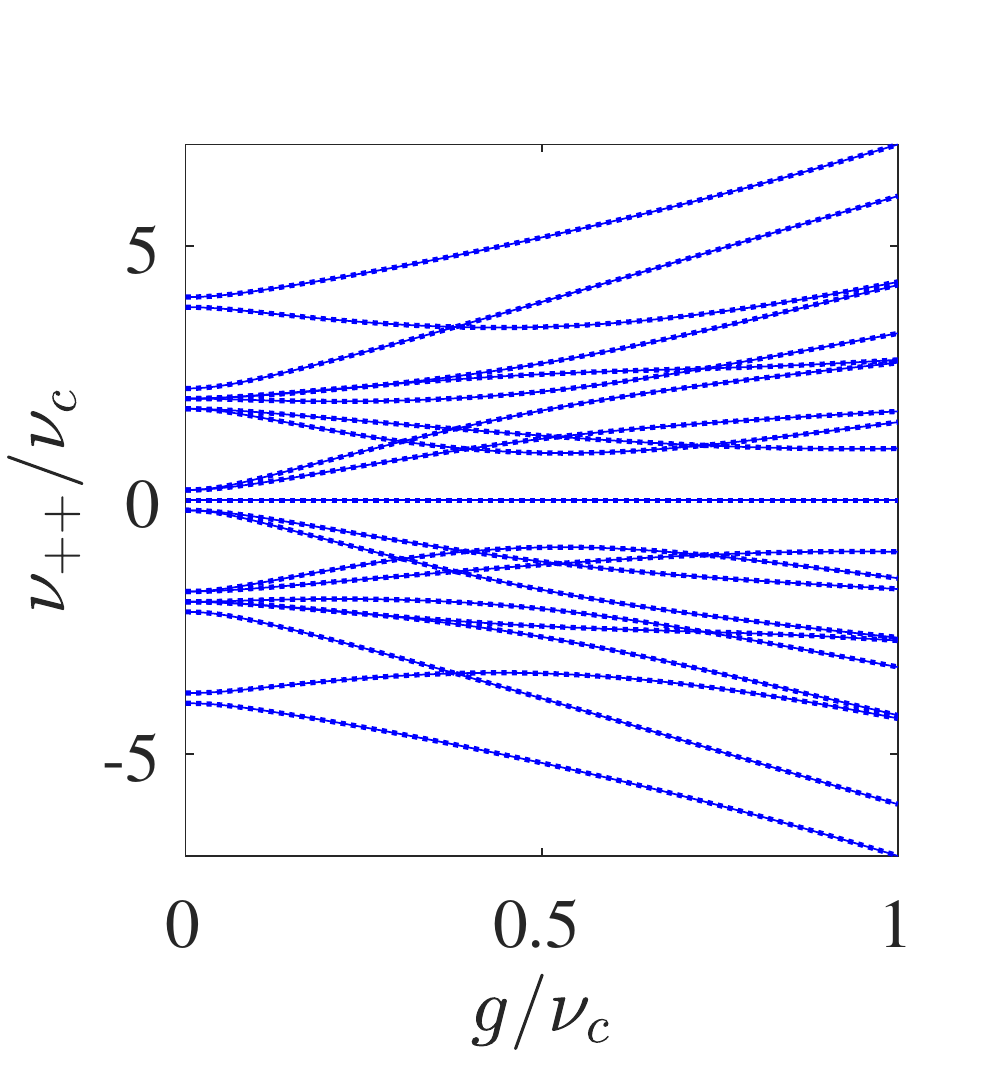}%
} \hfill
\subfloat[\label{subfig:QuOpRabiSpecPM}]{%
\includegraphics[scale=0.40]{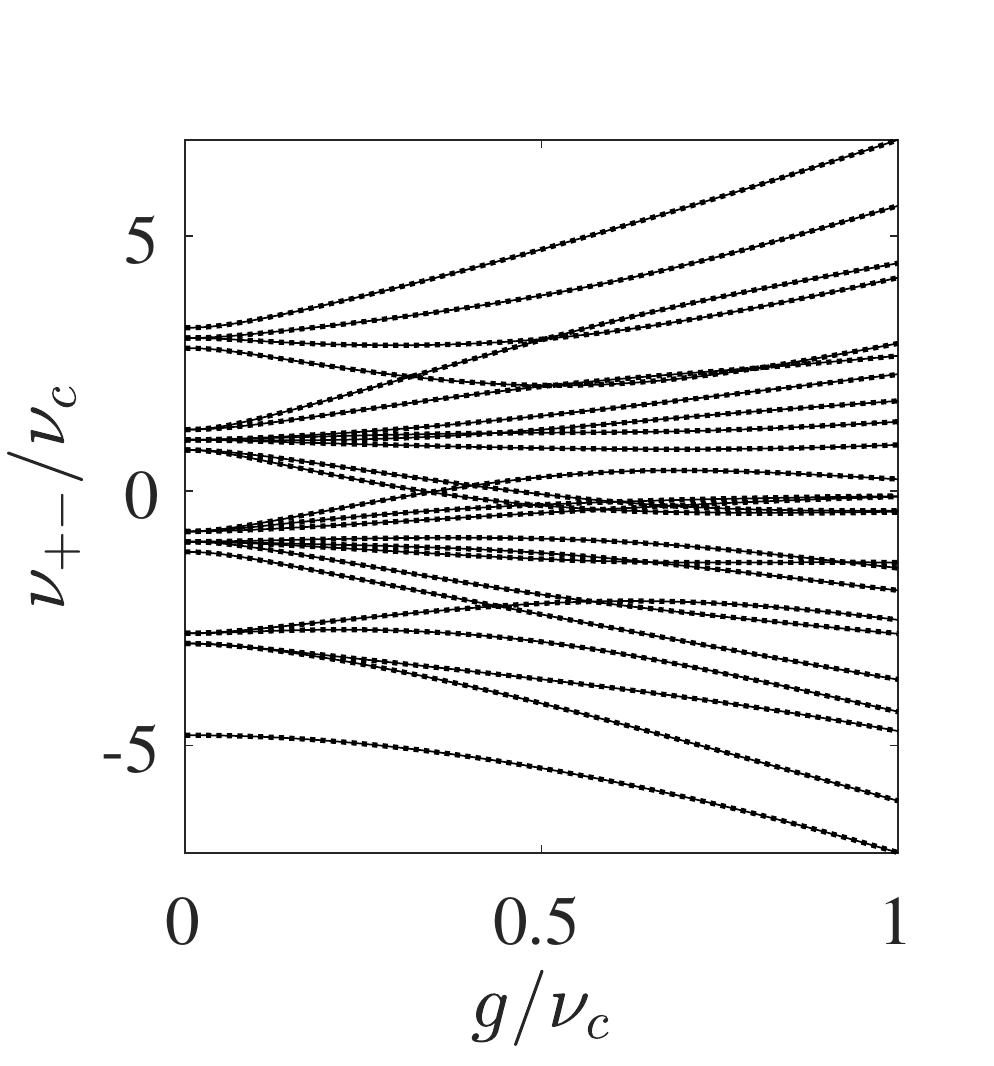}%
}\\
\subfloat[\label{subfig:QuOpRabiSpecMP}]{%
\includegraphics[scale=0.40]{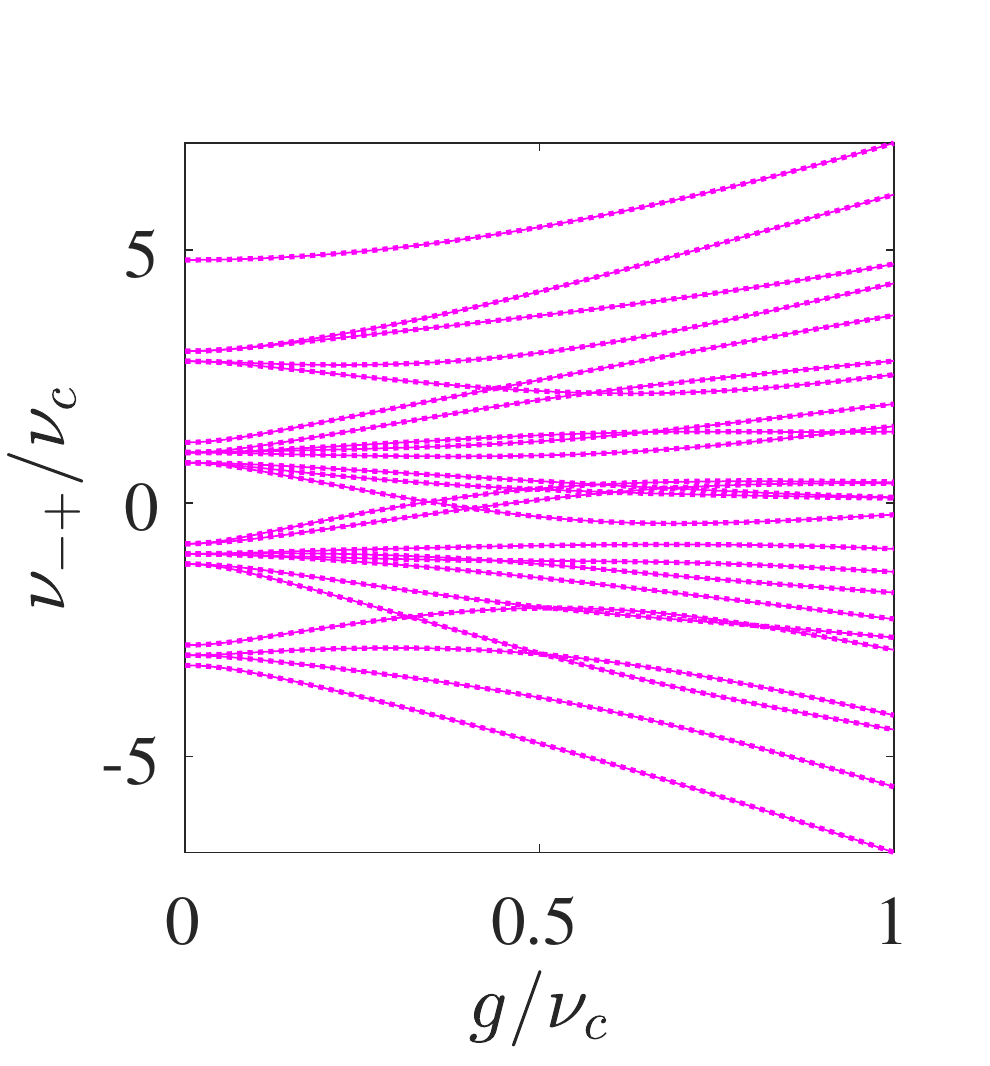}%
}\hfill
\subfloat[\label{subfig:QuOpRabiSpecMM}]{%
\includegraphics[scale=0.40]{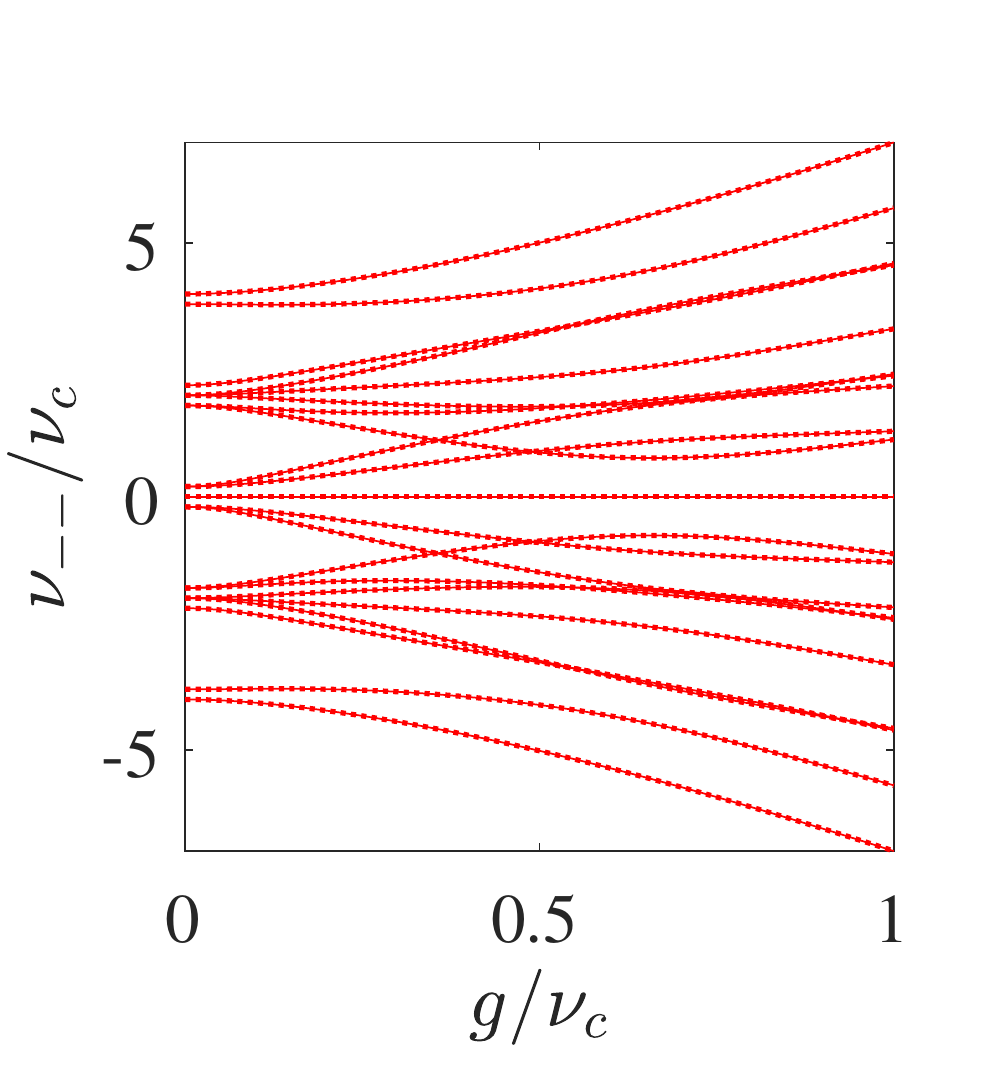}%
}
\caption{Quantum open Rabi eigenfrequencies for the same parameters as in Fig.~2 of the main paper as a function of light-matter coupling $g$. The eigenvalues are denoted by four colors. \textcolor{blue}{Blue} is for the $\color{blue}\ket{p_s=+}\ket{p_a=+}\color{black}$ subspace of the full effective Hamiltonian, \textcolor{red}{red} for $\color{red}\ket{p_s=-}\ket{p_a=-}\color{black}$, black for $\ket{p_s=+}\ket{p_a=-}$ and \textcolor{byzantine}{purple} for $\color{byzantine}\ket{p_s=-}\ket{p_a=+}$. The solid lines show the result for the full effective Hamiltonian, while the dotted line is the phenomenological case. The Hilbert space cutoff for each boson, i.e. $\hat{b}$ and $\hat{B}$, has been chosen as $N_c=4$ for clarity.}
\label{fig:QuOpRabiSpec} 
\end{figure}
\begin{figure}[t!]
\centering
\subfloat[\label{subfig:QuOpRabiDissPP}]{%
\includegraphics[scale=0.40]{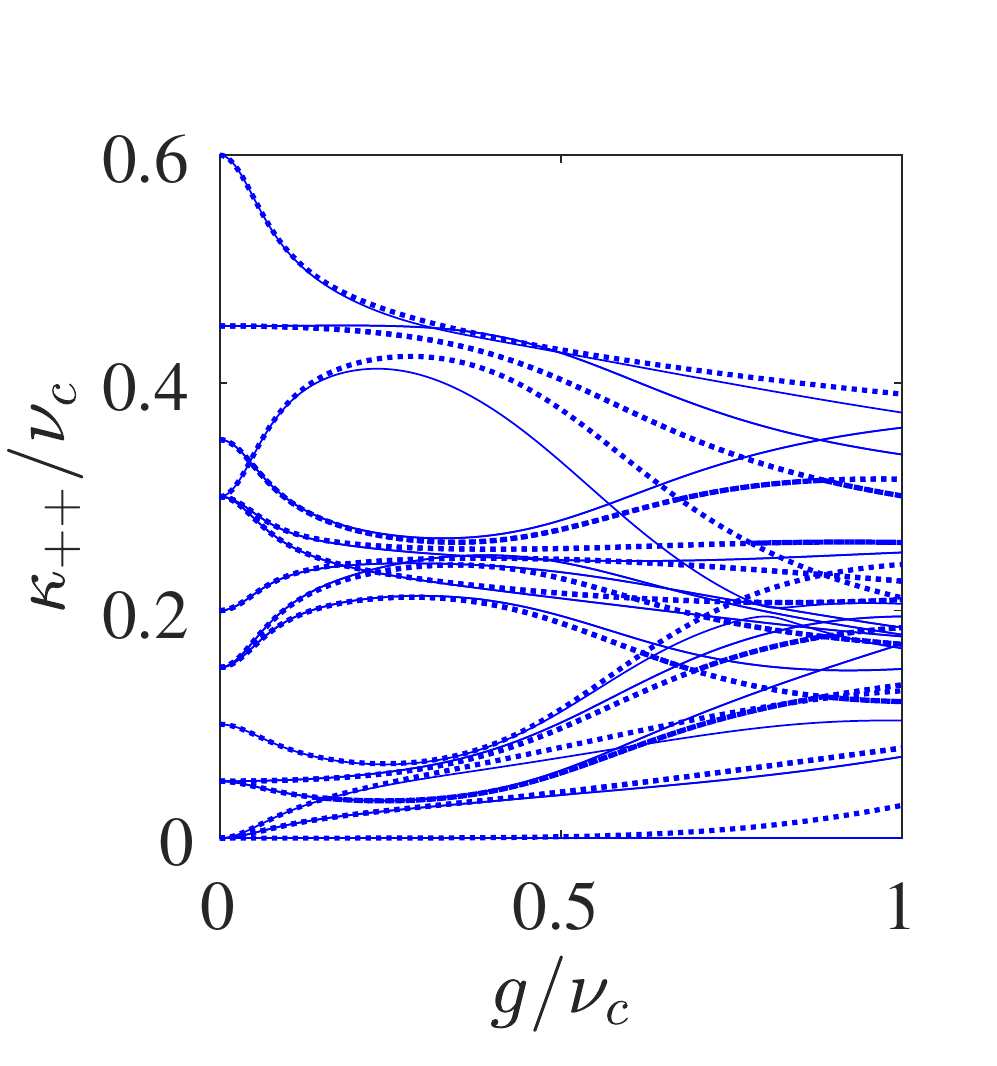}%
} \hfill
\subfloat[\label{subfig:QuOpRabiDissPP}]{%
\includegraphics[scale=0.40]{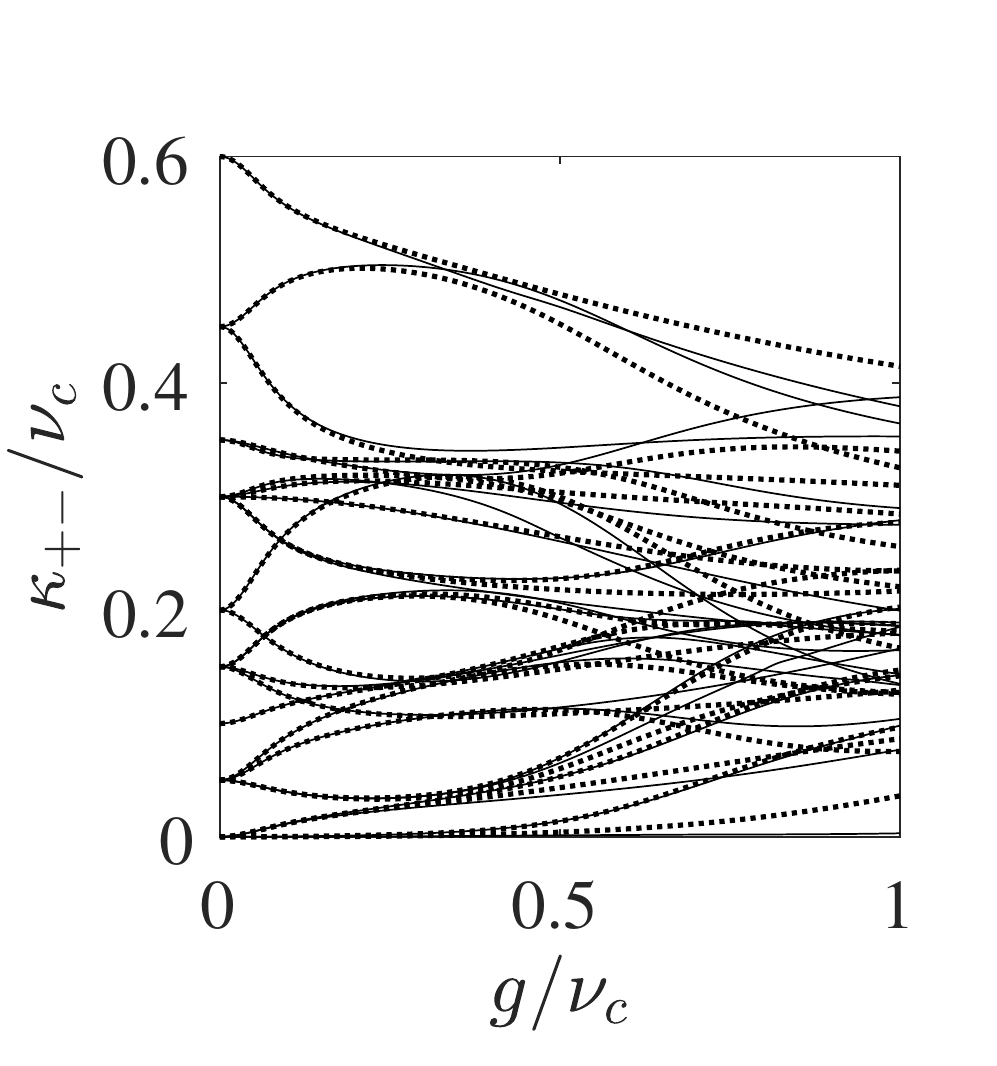}%
}\\
\subfloat[\label{subfig:QuOpRabiDissPP}]{%
\includegraphics[scale=0.40]{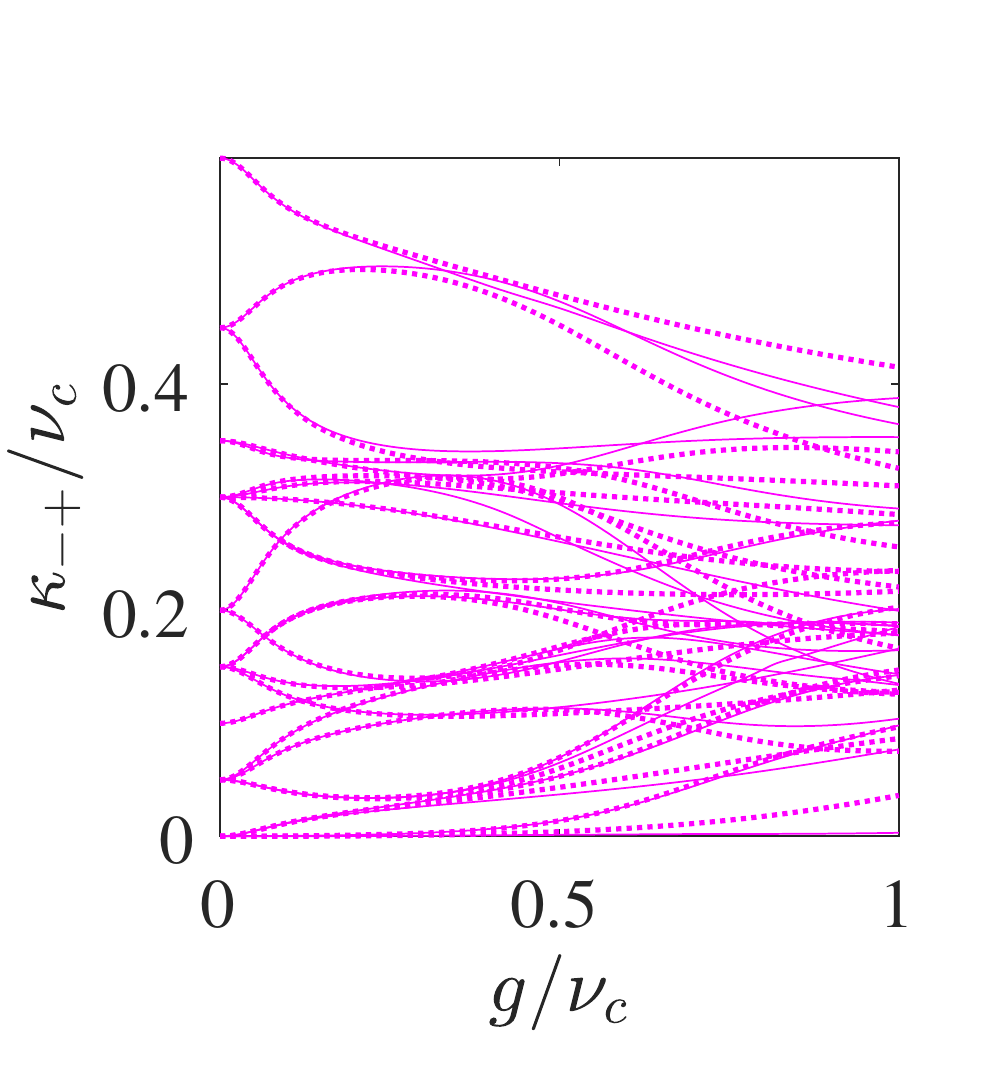}%
}\hfill
\subfloat[\label{subfig:QuOpRabiDissPP}]{%
\includegraphics[scale=0.40]{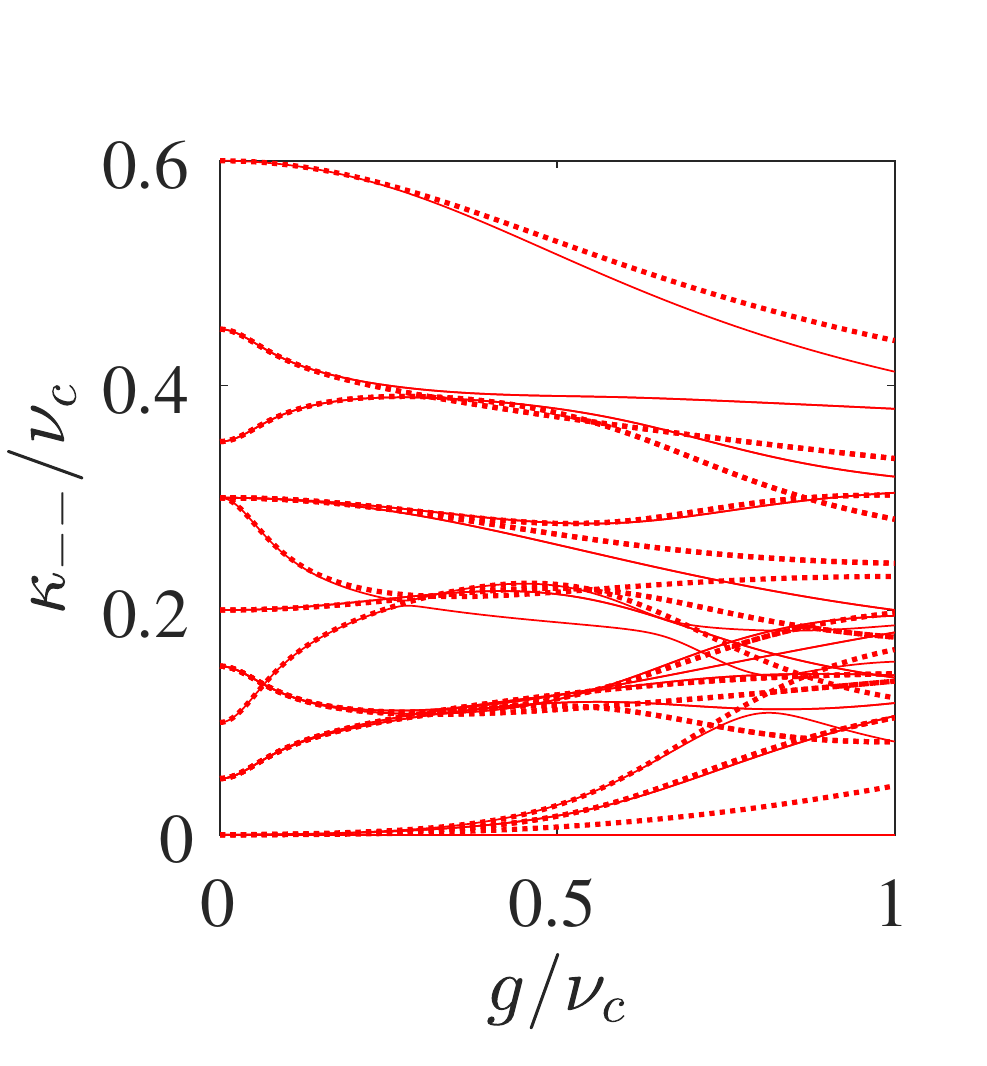}%
}
\caption{Quantum open Rabi eigendecays for the same parameters and plotting conventions as in Fig.~\ref{fig:QuOpRabiSpec} as a function of light-matter coupling $g$.}
\label{fig:QuOpRabiDiss} 
\end{figure}
  
In this subsection, we discuss the derivation of the full effective Hamiltonian approach for the quantum Rabi model with two-photon relaxation and check the validity of the phenomenological results for the spectrum. We start by the Lindblad equation 
\begin{align}
\dot{\hat{\rho}}(t)=-i[\hat{H}_s,\hat{\rho}(t)]+2\kappa_{c2}\mathcal{D}[\hat{a}^2]\hat{\rho}(t),
\label{Eq:EfHamRabi-Lindblad Eq 1}
\end{align}
where we have included a two-photon relaxation for the cavity mode. 

Importantly, the two-photon dissipator respects the $Z_2$ symmetry of the Rabi model, since
\begin{align}
\hat{P}^{\dag}\mathcal{D}[\hat{a}^2]\hat{P}=\mathcal{D}[(-\hat{a})^2]=\mathcal{D}[\hat{a}^2].
\label{Eq:EfHamRabi-ParSym of D[a]}
\end{align}
Therefore, we expect that the full Lindblad equation become block-diagonal in the parity sector. To see this explicitly, we first divide the dissipator into decay and collapse contributions and rewrite the Lindblad Eq.~(\ref{Eq:EfHamRabi-Lindblad Eq 1}) as
\begin{align}
\dot{\hat{\rho}}(t)=-i\left[\hat{H}_{\text{s,ef}}\hat{\rho}(t)-\hat{\rho}(t)\hat{H}_{\text{s,ef}}^{\dag}\right]+2\kappa_{c2}\hat{a}^2\hat{\rho}(t)\left(\hat{a}^{\dag}\right)^2,
\label{Eq:EfHamRabi-Lindblad Eq 2}
\end{align}
where the effective system Hamiltonian reads
\begin{align}
\hat{H}_{\text{s,ef}}\equiv \frac{\nu_q}{2}\hat{\sigma}^z+\nu_c\hat{a}^{\dag}\hat{a}+g\left(\hat{a}+\hat{a}^{\dag}\right)\hat{\sigma}^x-i\kappa_{c2}(\hat{a}^{\dag})^2\hat{a}^2.
\label{Eq:EfHamRabi-H_R,ef}
\end{align}
Second, we reexpress the Lindblad Eq.~(\ref{Eq:EfHamRabi-Lindblad Eq 2}) in the number-parity basis. Using definition~(\ref{Eq:ParSym-Def b}) we find that 
\begin{align}
\hat{b}^2=(\hat{\sigma}_x\hat{a})^2=\hat{\sigma}_x^2\hat{a}^2=\hat{a}^2,
\label{Eq:EfHamRabi-Sol for H_s,ef}
\end{align} 
from which we can reexpress the effective system Hamiltonian~(\ref{Eq:EfHamRabi-H_R,ef}) as
\begin{align}
\hat{H}_{\text{s,ef}}=\nu_c\hat{b}^{\dag}\hat{b}-\frac{\nu_q}{2}(-1)^{\hat{b}^{\dag}\hat{b}}\hat{P}_s+g\left(\hat{b}+\hat{b}^{\dag}\right)-i\kappa_{c2}(\hat{b}^{\dag})^2\hat{b}^2.
\label{Eq:EfHamRabi-Sol for H_s,ef}
\end{align}
The Lindblad Eq.~(\ref{Eq:EfHamRabi-Lindblad Eq 2}) then takes the form
\begin{align}
\dot{\hat{\rho}}(t)=-i\left[\hat{H}_{\text{s,ef}}\hat{\rho}(t)-\hat{\rho}(t)\hat{H}_{\text{s,ef}}^{\dag}\right]+2\kappa_{c2}\hat{b}^2\hat{\rho}(t)(\hat{b}^{\dag})^2=0.
\label{Eq:EfHamRabi-Lindblad Eq 3}
\end{align}

Comparing Eq.~(\ref{Eq:EfHamRabi-Lindblad Eq 3}) to the generic Lindblad form~(\ref{Eq:EffHam-Lindblad 2}) and following the discussion in Sec.~(\ref{SubApp:GenDis}), we can define a full effective Hamiltonian as
\begin{subequations}
\begin{align}
\hat{H}_{\text{u,ef}}=\hat{H}_{\text{s,ef}}-\hat{H}_{\text{a,ef}}+2i\kappa_{c2}\hat{b}^2\hat{B}^2,
\label{Eq:EfHamRabi-Def of H_u,ef}
\end{align}
where $\hat{H}_{\text{s,ef}}$ and $\hat{H}_{\text{a,ef}}$ are defined as
\begin{align}
\begin{split}
\hat{H}_{\text{s,ef}}&=\nu_c\hat{b}^{\dag}\hat{b}-\frac{\nu_q}{2}e^{i\pi\hat{b}^{\dag}\hat{b}}\hat{P}_s\\
&+g\left(\hat{b}+\hat{b}^{\dag}\right)-i\kappa_{c2}(\hat{b}^{\dag})^2\hat{b}^2,
\end{split}
\label{Eq:EfHamRabi-Sol for H_s,ef}\\
\begin{split}
\hat{H}_{\text{a,ef}}&=\nu_c\hat{B}^{\dag}\hat{B}-\frac{\nu_q}{2}e^{i\pi\hat{B}^{\dag}\hat{B}}\hat{P}_a\\
&+g\left(\hat{B}+\hat{B}^{\dag}\right)+i\kappa_{c2}(\hat{B}^{\dag})^2\hat{B}^2.
\end{split}
\label{Eq:EfHamRabi-Sol for H_a,ef}
\end{align}
\end{subequations}
In Eqs.~(\ref{Eq:EfHamRabi-Def of H_u,ef}-\ref{Eq:EfHamRabi-Sol for H_a,ef}), $\hat{b}$ and $\hat{P}_s$ denote the cavity annihilation and parity operators for the system sector, while $\hat{B}$ and $\hat{P}_a$ are the counterparts for the auxiliary sector. 

Even though it is in principle possible to obtain recursion relations for the eigenfrequencies and eigenmodes of the full effective Hamiltonian~(\ref{Eq:EfHamRabi-Def of H_u,ef}), due to the larger Hilbert space compared to the phenomenological treatment, the results will be more involved. Alternatively, we avoid analytics and compute the complex spectrum of the full effective Hamiltonian~(\ref{Eq:EfHamRabi-Def of H_u,ef}) numerically. 

To make a meaningful comparison to the phenomenological result of Sec.~\ref{App:ClRabiSpec}, we can in principle follow two possible routes. The first possibility is to employ spectral decomposition of the full spectrum, which provides a reduced spectrum that could be in principle compared with the phenomenological result. The Rabi model is not quadratic, and a numerical study shows that the aforementioned spectral decomposition [Eq.~(\ref{Eq:EffHam-TRD for Om})] does not hold for this system. Therefore, we follow the second possibility of using the Hamiltonian $\hat{H}_{\text{u,ph}}=\hat{H}_{\text{s,ef}}-\hat{H}_{\text{a,ef}}$, that lacks the collapse terms, and compare its spectrum with that of $\hat{H}_{\text{u,ef}}$ in Eq.~(\ref{Eq:EfHamRabi-Def of H_u,ef}). An example for the spectrum is shown in Figs.~\ref{fig:QuOpRabiSpec} and~\ref{fig:QuOpRabiDiss} for the real part (frequency) and imaginary part (decay) of the complex spectrum, respectively. For clarity, the spectrum has been partitioned into four (i.e. $p_s=\pm, p_a=\pm$) possible parity subspaces. Following the results in the main paper, we have kept the lowest 5 levels for the system and auxiliary bosons. Therefore, we observe $5^2=25$ distinct quantum levels for each of the four possible parity subspaces. Note that the collapse term has a representation in the number basis that is lower triangular. Therefore, at $g=0$, we expect the spectrum of the two  Hamiltonians to be exactly the same. On the other hand, the coupling terms have tridiagonal representations and due to the distinct interplay of coupling with the dissipation, with and without the collapse, we expect to get deviations at larger values of $g$ as seen in Fig.~\ref{fig:QuOpRabiDiss}. The real frequencies, on the contrary, barely show any modification due to the collapse terms    (See Fig.~\ref{fig:QuOpRabiSpec}). 
\bibliography{RabiTwoPhotonBib}
\end{document}